# A Flexible Two-stage Pricing Strategy Considering Learning Effects and Word-of-Mouth

Yanrong Li, Lai Wei and Wei Jiang

Antai College of Economics and Management, Shanghai Jiao Tong University

**Abstract**

This paper proposes a flexible two-stage pricing strategy for nondurable (such as electronics) products, whose selling price is marked-down at a flexible time point during the lifecycle. We consider the learning effect of manufacturers, which is reflected by the decrease of average cost with the accumulation of production. Moreover, word-of-mouth (WOM) of existing customers is used to analyze future demand, since many customers make the purchase decision according to WOM. We theoretically prove the existence and uniqueness of the optimal switching time between the two stages and the optimal price in each stage. Besides, warranty is another important factor of electronic products considered in this work. The interaction between warranty and word-of-mouth is also analyzed. Interestingly, our findings indicate that (1) the main reason for manufacturers to mark-down prices for electronic products pertains to the learning effects; (2) even though both internal factors (e.g., the learning effects of manufacturers) and external factors (e.g., the price elasticity of customers) have impacts on product price, their influence on manufacturer's profit is widely divergent; (3) warranty weakens the influence of external advertising on the reliability perception of customers because warranty price partially reflects the actual reliability of products; (4) and the optimal warranty can increase the profits for the manufacturer by approximately 10%.

# 1. Introduction

Manufacturers often make pricing decisions based on customers' characteristics and complex market environments (Liu et al., 2019). In general, a huge challenge for manufacturers is to determine appropriate prices for their products, to be recognized by customers, and capture a higher market share with more notable brand influence (Lei et al., 2017). For example, for electronic products, pricing decisions generally depend on the different stages in the lifecycle. According to Pan et al. (2009), for electronics, during their whole lifecycle from launching to being discontinued, a fixed pricing strategy cannot maximize the profit of the manufacturer. Therefore, a large amount of research focuses on dynamic pricing strategies during the whole lifecycle of the products (Andreas et al., 2012; Nair and Closs, 2006).

In general, product pricing depends on many internal and external factors. Internal factors source from manufacturers. Two main internal factors are product reliability and the learning effects of the manufacturer. It is often challenging for manufacturers to control the reliability of each item; nevertheless, product reliability is a significant factor influencing customers' decisions. In addition, with the development of technology and proficiency of production, learning effects reflect the manufacturing experience of manufacturers, and as a result, the average production cost of each product is decreased. Therefore, Cheng et al. (2019) claimed that learning effects must be considered in the operational process, especially for electronic products.

External factors source from customers. Customers' price elasticity and word-of-mouth (WOM) are typical and important external factors. Particularly, WOM changes with the accumulation of sales volume (Duan et al., 2008). Generally, positive WOM increases the sales volume of products. In practice, the demand of customers cannot be determined by the price and WOM. Therefore, a stochastic demand during products' lifecycle is necessary to be discussed.

In general, manufacturers strive to maximize a product's lifecycle profit through the pricing decision. Since manufacturers cannot control sales volume, ensuring a reasonable price is an effective way for them to maximize the profit (De and Zaccour, 2019; Chen et al., 2012). Therefore, establishing a pricing decision considering both internal and external factors is necessary.

Nevertheless, for certain nondurable products in practice, such as most consumer electronics, prices do not change frequently. Instead, prices of such products remaining steady for a certain period and then drop to a lower level [1]. Based on this phenomenon, our work aims to explain the reason behind the change of price and proposes the optimal pricing strategy for manufacturers considering both internal (e.g., learning effect) and external factors (e.g., WOM). Beforehand, several questions are put forward and to be solved in our work:

- Among learning effect and WOM, which is the key reason for the price change during the product lifecycle?

---

[1] From https://tool168.cn/ and http://detail.zol.com.cn/ which are price monitoring platforms, the history price of products in different e-commerce websites. The price of electronic products always decreases.

- How to determine the optimal price in each stage and the switching time of the price considering the WOM?
- Between the internal and external factors, which one is more important for the manufacturer's total profits.

To answer these questions and for analytical tractability, we propose a flexible two-stage model in continuous time considering both internal and external factors to maximize the total profit of the manufacturer. Our model is flexible because the price switching time is a decision variable within the continuous sales horizon. It brings more difficulties to find the optimal solution than the normal two-stage model. Therefore, the innovations in our paper are (1) the flexible two-stage pricing model is established; (2) we obtain a closed-form solution for the model; (3) the uniqueness of the optimal prices in two stages and price switching time is also proved. Furthermore, the effects of internal factors source from the manufacturer and external factors source from customers are also discussed, respectively. By analyzing the effects of WOM, we discover that an underestimation of perceived reliability hurts profitability whereas an overestimation may not bring benefit to the manufacturer.

We further consider warranty service in the above two-stage model to investigate its interaction with WOM. Since warranty price often reflects product reliability and also influences customers' WOM, we incorporate warranty in our flexible two-stage model and derive the optimal warranty price. We find that warranty price can provide information on actual product reliability and reduce the gap between overestimation and underestimation of perceived reliability in WOM. Moreover, warranty service brings approximate 10% increments of the total profit.

The remainder of this paper is organized as follows. Section 2 provides a literature review related to our work. Section 3 describes the model considering the learning effect of the manufacturer and WOM of customers and derives the optimal prices and price switching time. In Section 4, an extended model considering the interaction between warranty and WOM is described. Moreover, the numerical analysis presented the effects of warranty in the model. Finally, our findings with managerial insights, as well as discussions of future research are summarized in Section 5. All the proofs and the robustness verification in our model are presented in the Appendix.

## 2. Literature review

Several research branches are related to this work. We summarize these branches into four categories and describe the key results in each category.

**2.1 Previous research on discount pricing strategy**

One of the most key research domains is to identify a proper pricing strategy for manufactures to gain more profits. As one of the most important key factors, the customers' strategy has been extensively considered in recent research.

In the context of customers with different strategies, Aviv and Pazgal (2008) proposed a stylized model to formulate the optimal pricing strategy of a product with forward-looking (strategic) consumers. In the research setting, the seller announced his/her pricing strategy at first, and the customers acted as followers in a Stackelberg scenario, thereby formulating their purchasing behaviors. Su and Zhang (2008) also considered the customers' strategic decisions based on the customers' anticipations of selecting the purchase time to maximize their surplus. Considering a similar scenario, Cachon and Swinney (2009) proposed a more detailed model, which divided consumers into three more complex categories, namely, myopic consumers, bargain-hunting consumers, and strategic consumers. Li and Huh (2012) focused on the pricing decision for short life-cycle products when the customers' price-sensitivity is changed, and they got the optimal set of prices and price switching time. Papanastasiou and Savva (2017) discussed a preannounced pricing policy in a simple two-period case and formulated an optimal pricing strategy considering social learning. Dai and Nu (2020) focused more on the emerging sharing modes (such as consumer-to-consumer and business-to-consumer) and examined the equilibrium pricing considering the limited capacity of manufacturers. Nosrat et al. (2021) proposed a pricing model with network effects, which are described by the choice model in customers' demand.

In this work, different from various extensions and applications of the newsvendor model (Khouja, 1995;1999), we focus on electronic products, whose pricing markdown is different from the salvage value in the basic newsvendor model. Moreover, the demand for the second period also depends on the switching time of the two periods.

**2.2 Previous research on learning effects**

The second stream of the related studies pertains to learning effects. In most manufacturing processes, the learning effect means that productivity improves over time, and as a result, the production time becomes shorter and/or production cost reduces (Mosheiov, 2001a). Traditionally, learning effects are widely considered in scheduling optimizations for both single machine and parallel machine (Cheng and Wang, 2000; Mosheiov, 2001b). In recent years, many research jointly considered the learning effects and other production factors simultaneously. Mehdizadeh et al. (2018) proposed the production planning model based on learning effects and machine deterioration. As a result, they obtained the optimal levels of production rates, inventory and shortage, worker's hiring, and quantities of the products that are subcontracted.

With the general acceptance of the learning effect, many research took the learning effect as a key factor and considered it into their models for pricing strategy design. For example, Pei-Chun (2008) provided a model to analyze the optimal trajectories for pricing, quality, and production rate. Ke et al. (2011) considered a dynamic pricing strategy considering the cost learning effect, and they found that the learning effect is beneficial for manufacturers in long-term profitability but detrimental to the channel inefficiency. Crapis et al. (2016) analyzed the social learning mechanism and its effect on the seller's pricing decision. Considering an intuitive and non-Bayesian decision rule, they showed that consumers eventually learn the real product's quality. In our work, the learning effect is also considered as an important factor, and we combine it with WOM and warranty in the pricing strategy decision.

**2.3 Previous research on word-of-mouth (WOM)**

The third research stream is related to WOM. WOM marketing is treated as a form of free advertising for a product or service. WOM is shared by existing customers through their daily dialogues or the internet and is triggered by customer experiences. In many studies, WOM has been recognized as an essential factor in designing marketing strategies (Buttle 1998, Kumar et al. 2007). The related research pertains to consumer psychology, advertising strategy, and information communication.

A pioneer of WOM research was Arndt (1967), who first proposed the formal definition of WOM as a person-to-person communication of commercial entities. Early in the 1980s, the influence of WOM is considered in quality control and sales process (Lee, and Tapiero, 1986). Based on the early comprehension of WOM, Nyilasy (2005) summarized the WOM literature into four research areas, focused on why do people listen (Murray, 1991; Wirtz and Chew, 2002), what makes people talk (Derbaix and Vanhamme, 2003), what happens to the communicator after the WOM event (Dichter, 1966), and the power of WOM (Bone, 1995).

With the increasing popularity of the Internet, Hennig-Thurau et al. (2004) first extended the traditional WOM to e-WOM, which corresponded to "any positive or negative statement regarding products or services, transmitted via the Internet". Trusov et al. (2009) compared the effects of WOM versus traditional marketing based on the analysis of an internet social networking site. In an empirical study, Chevalier and Mayzlin (2006) noted that WOM positively affected consumers' purchasing behavior at two internet retail sites.

In addition to empirical studies, many researchers proposed several models considering WOM. Yu, et al. (2016) noticed the influence of customer reviews and considered these reviews to make pricing decisions. The authors used a Bayesian decision model to describe how the customers evaluated the quality through WOM. Godes (2017) studied how to balance quality and WOM in different cases. Kamada and Öry (2020) examined the use of referral rewards and a free contract to characterize the optimal incentive plan for WOM. Our work considers the effects of WOM and incorporates it as a key factor that affects the sales volume and pricing strategy.

**2.4 Previous research on the warranty**

The last stream of the related studies pertains to the warranty. The warranty refers to the obligation of manufacturers to repair or replace a failed product within a certain promissory period. Unlike the basic warranty, which is bundled with the product, an extended warranty is an option for the consumers to buy. Considerable research has focused on the extended warranty policy design (Gallego et al., 2015; Ammar and Surendra, 2017; Wang et al., 2020).

Besides, many scholars considered extended warranty service as a factor in the production and pricing processes. Hartman and Laksana (2009) designed the pricing menus of extended warranty based on different consumer's risk characterizations. Darghouth et al. (2017) jointly optimized product design, selling price, and warranty period. Liu et al. (2020) considered optimal pricing and production strategies under a free replacement warranty strategy in a two-period setting. Here a life-long extended warranty is considered as an influential factor of the reliability estimation by the customers. Different from previous research, we jointly optimize the selling price, its switching time, and (extended) warranty price for manufacturers.

## 3. Optimal pricing strategy with WOM

This paper considers a pricing problem for a monopolistic manufacturer, who sells nondurable products such as electronics to consumers. Nondurable products exhibit certain characteristics such as short lifecycle, frequent generation updates, and discrete price changes. Thus, the pricing strategy of these products is different from the hedged or durable products. Generally, because of the popularization of manufacturing techniques, the price of electronic products is relatively high during the early phase after launch and decreases with the length of the sales period. However, if price changes too frequently, customers might feel unfair and react negatively (Rotemberg, 2005). To avoid this, in practice the change of price does not occur continuously (usually in a stepwise form). Therefore, a critical issue for the manufacturer is to determine the price in each stage and the time point to switch prices.

Furthermore, because WOM significantly affects the consumers' perception, especially for new products (Campbell, 2015, Hervas-Drane, 2015), we consider the WOM in our model to make pricing decisions for new nondurable products. To simplify this multistage pricing problem, we take two-stage as an example to conduct theoretical analysis, that is to say, a two-stage pricing optimization problem during a finite horizon is discussed from the perspective of the manufacturer. The sales period is normalized to $[0, 1]$, where 0 and 1 represent the product launch time and the end of the sales horizon, respectively. In the flexible two-stage problem, we define the flexible switching time between two stages as $\theta \in (0, 1)$, and the selling prices $p_1$ and $p_2$ before and after the switching time $\theta$, respectively. Because demand depends highly on prices, identifying the optimal prices and switching time is important for a monopolistic manufacturer.

## 3.1 Model set-up of demand and profit

In this subsection, the demand of customers and total lifecycle profit of the manufacturer are introduced. Demand is always a stochastic function over price and sales period, as well as WOM. Therefore, to define the demand function, we first introduce the effects of these three factors.

For the effect of price, the existing works widely employed log-linear demand functions to describe it. Because the stable demand elasticity is consistent with the strategy of a monopolistic manufacturer. Traditionally, a log-linear model was used to estimate the demand functions for irreplaceable energy resources such as water (Gunatilake et al., 2001; Schleich and Hillenbrand, 2009) and electricity (Fan and Hyndman, 2011). In recent research, log-linear functions have been applied in the service industry (Chu, 2017) and the retail industry (Hekimoğlu et al., 2019). Following the literature, we define the sales volume at stage $i$ ($i \in \{1,2\}$) as $Q_i$, based on the log-linear demand function.

Generally, the demand in different stages of products' lifecycle is always changed, Hence, by defining $\lambda(t)$ as the probability density function of demand distribution over sales time, we describe the demand change over the sales horizon. As described in Sections 3 and 4, the theoretical analysis is performed based on the assumption that $\lambda(t)$ follows a uniform distribution. This assumption is relaxed in Appendix I, which proves the robustness of our conclusions. Moreover, we use $\gamma$ to denote the exponential coefficient of the demand elasticity, which represents the influence of price on sales volumes. Therefore, based on the log-linear demand model (Hekimoğlu et al. 2019), we have

$$\begin{cases} Q_1^0(p_1, \theta) = \frac{1}{p_1^\gamma} \int_0^\theta \lambda(t) dt \\ Q_2^0(p_2, \theta) = \frac{1}{p_2^\gamma} \int_\theta^1 \lambda(t) dt, \end{cases} \quad (3.1)$$

where $Q_i^0$ ($i \in \{1,2\}$) denotes the basic demand only taking price into account.

According to Lei et al. (2017), WOM is represented by the customers' perceived failure rate or product reliability and reflected in the demand $Q_i^0$ in Equation (3.1). The customers' perceived reliability depends on product quality and customers' initial perception, which is a time-varying variable according to the usage of products. According to the empirical study by Duan et al. (2008), positive WOM can stimulate and increase sales volume. Compared with the basic demand in Equation (3.1), the WOM model exhibits differences in terms of sales volume. The fluctuation of sales volume owes to the difference in actual reliability. Thus, the difference is expressed as $r_\Delta(t) = r_c(t) - r_m$, where $r_c(t)$ is the customers' perceived reliability at time $t$, and $r_m$ is actual reliability, which is a factor unknown to customers. Therefore, the demand functions influenced by WOM in the two sales stages are

$$\begin{cases} Q_1(p_1, \theta) = \frac{1}{p_1^\gamma} \int_0^\theta \lambda(t)(1 + r_\Delta(t)) dt \\ Q_2(p_2, \theta) = \frac{1}{p_2^\gamma} \int_\theta^1 \lambda(t)(1 + r_\Delta(t)) dt, \end{cases} \quad (3.2)$$

The customers' perceived reliability is based on historical WOM. According to Campbell (2015) and Lei et al. (2017), we divide the sales time into $N$ periods. In the $j^{th}$ ($0 \leq j \leq N$) period, customers' perceived reliability is defined as

$$r_c(j) = \begin{cases} r_0 & \text{for } j = 1 \\ \beta r_c(j-1) + (1-\beta) r_m & \text{for } j = 2, \dots, N, \end{cases} \quad (3.3)$$

where $r_0$ is customers' initial perception of product reliability before product launch, which is influenced by the external factors such as the advertising and manufacturers' brand effects, and $\beta$ is a smooth factor that ranges from 0 to 1. A larger $\beta$ corresponds to a more historical WOM and a smaller amount of actual reliability learned by customers in each period.

Given the price switching time from $p_1$ to $p_2$ occurs in the $M^{th}$ period ($M \in (0, N)$ and equivalently $\theta = \frac{M}{N}$), by substituting Equation (3.3) into demand Equation (3.2), we have the demand functions in the two sales stages with WOM as

$$\begin{cases} Q_1(p_1, M) = \frac{1}{p_1^{\gamma}} \left[ \sum_{i=1/N}^{M/N} \int_{i-\frac{1}{N}}^{i} \lambda(t)(1 + r_c(i \cdot N) - r_m) dt \right] \\ Q_2(p_2, M) = \frac{1}{p_2^{\gamma}} \left[ \sum_{i=\frac{M+1}{N}}^{1} \int_{i-\frac{1}{N}}^{i} \lambda(t)(1 + r_c(i \cdot N) - r_m) dt \right]. \end{cases} \quad (3.4)$$

With the accumulation of production time and products' quantity, the manufacturer becomes familiar with the production process. Consequently, operational cost decreases in terms of cumulative time or production volume (Feng and Chan., 2019), and the average manufacturing cost for each product decreases with time (Womer and Patterson, 1983; Li and Rajagopalan, 1998). This phenomenon occurs widely in the case of consumer electronics and reflects the learning effect of manufacturers. According to Arrow (1962) and Teng and Thompson (1983, 1996), manufacturers' learning effects might notably influence manufacturing cost and total profits. Therefore, in this work, the learning effects of manufacturers are incorporated in the model. According to Pourakbar et al. (2012), we use $c_m(t) = c \cdot e^{-\alpha t}$ to denote the function of manufacturing cost over time, where $c$ is the average manufacturing cost at time 0, and $\alpha > 0$ is the cost erosion factor per time of manufacturers. In particular, if $\alpha = 0$, learning effects do not exist, and the manufacturing cost is constant over time.

Based on the expressions of demand and manufacturing cost, the manufacturer can maximize its total profits by deciding the optimal selling price in the two sales stages ($p_1$ and $p_2$) and the switching time between these two stages ($\theta$). The manufacturer's profits in the two stages can be expressed as follows:

$$\begin{cases} \pi_1(p_1, M) = \frac{1}{p_1^{\gamma}} \left[ \sum_{i=1/N}^{M/N} \int_{i-\frac{1}{N}}^{i} \lambda(t)(1 + r_c(i \cdot N) - r_m(t))(p_1 - c \cdot e^{-\alpha t}) dt \right] \\ \pi_2(p_2, M) = \frac{1}{p_2^{\gamma}} \left[ \sum_{i=\frac{M+1}{N}}^{1} \int_{i-\frac{1}{N}}^{i} \lambda(t)(1 + r_c(i \cdot N) - r_m(t))(p_2 - c \cdot e^{-\alpha t}) dt \right]. \end{cases}$$

(3.5)

Moreover, the total profit over the two stages is $\pi(p_1, p_2, M) = \pi_1(p_1, M) + \pi_2(p_2, M)$. The aim is to find optimal values of $p_1$, $p_2$ and $\theta$ ($\theta = \frac{M}{N}$) that maximize the total profit. For convenience, we summarize notations of different parameters in Table 1.

**Table 1** Notations of parameters

| Parameters | Notations |
|---|---|
| $p_1$ | The selling price of products in stage 1 |
| $p_2$ | The selling price of products in stage 2 |
| $Q_1$ | Demand in stage 1 with WOM |
| $Q_2$ | Demand in stage 2 with WOM |
| $\pi_1$ | Total Profit in stage 1 with WOM |
| $\pi_2$ | Total Profit in stage 2 with WOM |
| $r_c$ | Customers' perceived reliability of products |
| $c_m$ | Manufacturing cost |
| $\alpha$ | Cost erosion factor due to manufacturer's learning effects |
| $\gamma$ | Exponential coefficient of the demand elasticity regarding price |
| $r_m$ | Real reliability of products |
| $\theta$ | Price switching time |

### 3.2 Optimal pricing strategy without WOM – a special case

To find the solution to the optimization problem in Equation (3.5), we start to analyze the optimality from a special case. In particular, when the customers' initial perceived reliability equals the actual reliability of the product, which means that WOM does not influence sales volume. In this case, the expressions of profits in Equation (3.5) are simplified as follows

$$\begin{cases} \pi_1(p_1, \theta) = \frac{1}{p_1^\gamma} \int_0^\theta \lambda(t)[p_1 - c_m(t)]dt \\ \pi_2(p_2, \theta) = \frac{1}{p_2^\gamma} \int_\theta^1 \lambda(t)[p_2 - c_m(t)]dt \end{cases} \quad (3.6)$$

To find the optimal values of $p_1$, $p_2$ and $\theta$ to maximize the total profit $\pi(p_1, p_2, \theta) = \pi_1(p_1, \theta) + \pi_2(p_2, \theta)$, and to prove the uniqueness of the solutions, we propose Theorems 1 and 2 in the following.

To start with, we determine the optimal $p_1$ and $p_2$ for a given $\theta$, as described in Theorem 1.

**Theorem 1.** *For a given $\theta = \theta_0$, by solving the optimization problem $\max_{p_1 \geq 0, p_2 \geq 0} \pi(p_1, p_2, \theta_0) = \max_{p_1 \geq 0} \pi_1(p_1, \theta_0) + \max_{p_2 \geq 0} \pi_2(p_2, \theta_0)$, we can obtain the* *optimal prices in two stages* $p_1^*(\theta_0) = \frac{c \cdot \gamma(1-e^{-\alpha\theta_0})}{\alpha\theta_0(\gamma-1)}$ *and* $p_2^*(\theta_0) = \frac{c \cdot \gamma(e^{-\alpha\theta_0}-e^{-\alpha})}{\alpha(1-\theta_0)(\gamma-1)}$. *Besides, we have* $p_1^*(\theta_0) > p_2^*(\theta_0)$ *when* $\alpha > 0$.

**Proof:** The proof is provided in Appendix B.

According to Theorem 1, we discuss the optimal prices in the general case where $\alpha > 0$ and the special case where $\alpha = 0$. Generally, $\alpha > 0$ corresponding to a positive learning effect of the manufacturer, the optimal price in the first stage is higher than that in the second stage. In contrast, in the special case where there is no learning effect of the manufacturer, i.e., $\alpha = 0$, we have $p_1^* = p_2^* = \frac{c \cdot \gamma}{\gamma-1}$ and the optimization problem degenerates to a single-stage pricing problem. Comparing these two cases, we note that when the manufacturer's learning effect exists, both the manufacturing cost and optimal prices in the two stages decrease simultaneously.

In practice, electronics price does not increase after launch. Neglecting the effects of the external marketing environment, the reason for the price mark-down in the second stage owes to the learning effects of manufacturers. Because the learning effects directly reduce the average manufacturing cost of products, higher-margin profits can be attained. Consequently, the manufacturer can lower the price to obtain a larger sales volume, and thus, higher profits.

After obtaining the optimal $p_1^*(\theta)$ and $p_2^*(\theta)$ for a given $\theta$ in Theorem 1, we substitute $p_1^*(\theta)$ and $p_2^*(\theta)$ into the profit function $\pi(p_1^*(\theta), p_2^*(\theta), \theta)$ to determine the optimal $\theta^*$. First, simplifying the total profit function in Equations (3.7) and (3.8), we have the following expression (the details are provided in Appendix C)

$$\pi(\theta) = \pi(p_1^*(\theta), p_2^*(\theta), \theta) = \frac{[\alpha(\gamma-1)]^{\gamma-1}}{\gamma^\gamma c^{\gamma-1}} \cdot \pi_0(\theta), \quad (3.7)$$

where

$$\pi_0(\theta) = \frac{\theta^\gamma}{(1-e^{-\alpha\theta})^{\gamma-1}} + \frac{(1-\theta)^\gamma}{(e^{-\alpha\theta}-e^{-\alpha})^{\gamma-1}}. \quad (3.8)$$

Subsequently, we need to prove that there exists a unique $\theta^* \in (0,1)$ to maximize the total profit $\pi(\theta)$. Because $\frac{[\alpha(\gamma-1)]^{\gamma-1}}{\gamma^\gamma c^{\gamma-1}}$ is a positive coefficient, we just analyze the monotonicity of $\pi_0(\theta)$, as indicated in Equation (3.8).

Based on the optimal $p_1^*(\theta)$ and $p_2^*(\theta)$ obtained in Theorem 1, we can examine the optimal solution of the price switch point ($\theta^*$) to maximize the total profit $\pi(p_1, p_2, \theta)$ in the following theorem.

**Theorem 2 (Uniqueness of optimal $\theta^*$):** *By substituting $p_1^*(\theta)$ and $p_2^*(\theta)$ into the profit function $\pi(p_1^*(\theta), p_2^*(\theta), \theta)$, we obtain the unique optimal $\theta^*$ to maximize the total profits. Finally, the unique optimal set of solutions for the optimization problem $\max_{p_1 \geq 0, p_2 \geq 0, \theta \in (0,1)} \pi(p_1, p_2, \theta)$ is $p_1^*(\theta^*)$, $p_2^*(\theta^*)$, and $\theta^*$.*

**Proof:** The proof is provided in Appendix D.

According to Theorems 1 and 2, we provide the closed-form solutions of the two-stage optimization problem and prove that the manufacturer has a unique set of optimal prices and the price switching time in the two stages. Apart from the decision strategy of the price and switching time based on Theorems 1 and 2, we also analyze the influence of different external parameters on the total profit and optimal solutions as follows.

**Observation 1:** *The optimal prices in the two stages decrease with the learning parameter $\alpha$ and price elasticity parameter $\gamma$, and the optimal price switching time $\theta$ increases with the two parameters.*

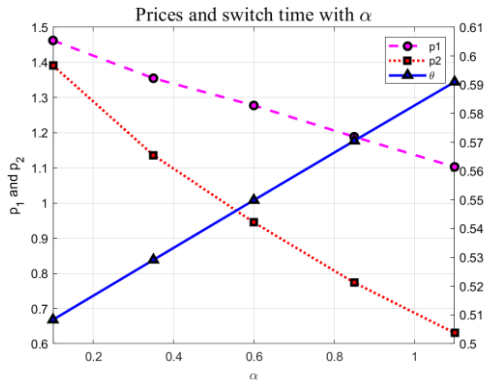 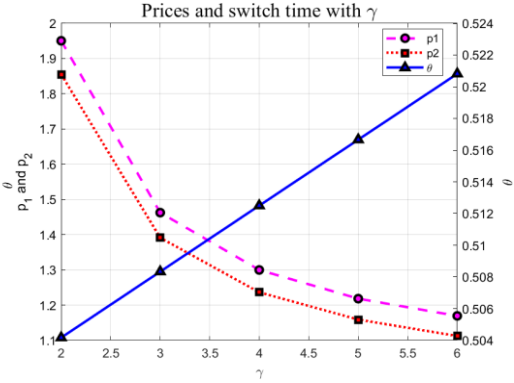

**Figure 1(a)**            **Figure 1(b)**

**Figure 1**   Prices and their switching time based on the learning effect and price elasticity parameter

A larger $\alpha$ corresponds to faster learning effects of the manufacturer and higher descent rate of the manufacturing cost ($c_m(t)$). Figure 1(a) illustrates that when $\alpha$ is large, the manufacturer tends to reduce the price to expand their sales volumes to achieve more profits. Moreover, the markdown value in the second stage ($p_1 - p_2$) also increases with the learning effects ($\alpha$). This phenomenon is consistent with the fact that the price erosion can be attributed to technology popularization and proficiency level increase. Moreover, the optimal price switching time ($\theta$) exhibits a positive relationship with the learning parameter ($\alpha$) and the markdown value in the second stage ($p_1 - p_2$). A lower price in the second stage corresponds to a delayed switching time (larger $\theta$) to buffer the loss on the price reduction. Similarly, for a lower initial price ($p_1$), the first stage is prolonged.

Figure 1(b) shows the influence of the price elasticity parameter ($\gamma$) on the optimal prices and switching time. For a larger $\gamma$, customers exhibit a more sensitive attitude toward the price, i.e. a slight change of price could result in a larger change in the sales volume. Therefore, with the increase in the price elasticity parameter ($\gamma$), the manufacturer tends to reduce the selling prices $p_1$ and $p_2$ to ensure high sales volumes. Since $p_1$ and $p_2$ decrease, the switching time is slightly delayed, to relieve the price reduction.

**Observation 2:** *The total profit decreases with the initial manufacturing cost at time 0 ($c$) and price elasticity parameter ($\gamma$), and the profit increases with the learning effect parameter ($\alpha$).*

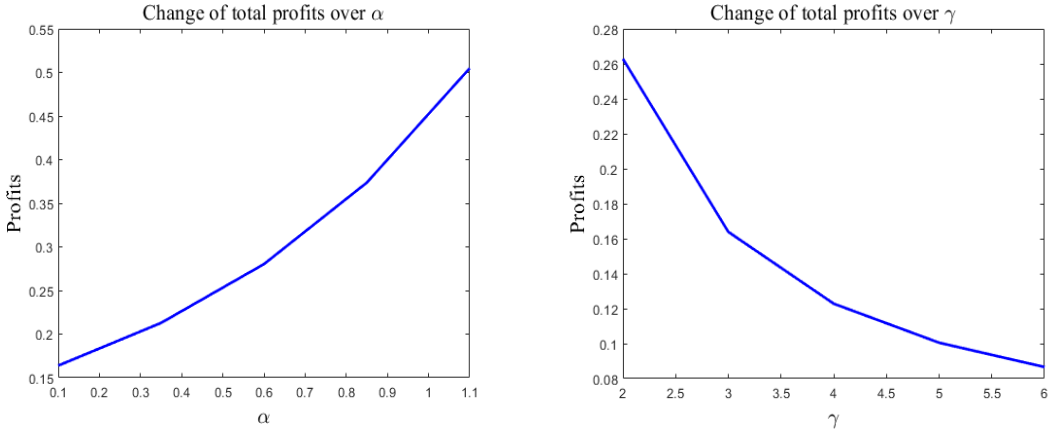

Figure 2(a)　　　　　　　　　　　　　Figure 2(b)

**Figure 2**　Total profits based on the learning effect and price parameters

The most important aim for a manufacturer is to ensure profits. In the context of the profits indicated in Equation (3.7), because ($\gamma - 1$) is positive, the total profit decreases with the initial manufacturing cost $c$, which means that a high initial cost reduces the total profits. To compare the prices with the initial manufacturing cost more conveniently, we normalize the initial manufacturing cost $c$ to 1.

According to Observation 1, the optimal $p_1$ and $p_2$ decrease with both the manufacturer's learning parameter $\alpha$ and the price elasticity parameter $\gamma$. In contrast to the prices, as shown in Figures 2(a) and 2(b), the total profit decreases with $\gamma$ and increases with $\alpha$. From the perspective of the manufacturer, the price reduction with $\alpha$ is proactive on account of the cost erosion, which provides the manufacturer an opportunity to lower the price and increase the sales and profits. In contrast, the elasticity parameter $\gamma$ is an external factor for the manufacturer, which represents the attitude of customers toward the prices. For a certain price, a larger $\gamma$ corresponds to lower demand. Moreover, a larger $\gamma$ indicates that a slight change in the price could lead to a sharp reduction in demand. Therefore, the manufacturer must control the prices at a low level carefully because there exists a substantial risk of losing the demand and profits.

**3.3 Optimal pricing strategy with WOM-general case**

In the general case with WOM, customers' initial perception of reliability is generally not equal to the actual reliability of the products. Recall that to model WOM, we discretize the overall sales process into $N$ periods, and the second stage ends in the $M^{th}$ period. Based on the expressions of WOM in Equation (3.3) and demand in Equation (3.4), we find the optimal prices to maximize total profit in Equation (3.5). Similar to the solution in Theorem 1, given a specific $M$ within $N$ stages, we can obtain the closed-form expressions of $p_1$ and $p_2$ presented in Equation (3.9).

$$\begin{cases} p_1\left(\frac{M}{N}\right) = \frac{c \cdot \gamma \cdot \sum_{i=1/N}^{M/N} e^{-\frac{\alpha i}{N}}(r_c(iN)+f_0)}{\alpha(\gamma-1)\sum_{i=1/N}^{M/N}(f_0+r_c(iN))} = \frac{c \cdot \gamma}{\alpha(\gamma-1)} \cdot \frac{N\left(1-e^{-\frac{\alpha M}{N}}\right)+(r_0-r_m)\frac{\left(1-\beta^M e^{-\frac{\alpha M}{N}}\right)}{1-\beta}}{M+(r_0-r_m)\frac{1-\beta^M}{1-\beta}} \\ p_2\left(\frac{M}{N}\right) = \frac{c \cdot \gamma \cdot \sum_{i=(M+1)/N}^{1} e^{-\frac{\alpha i}{N}}(r_c(iN)+f_0)}{(\gamma-1)\sum_{i=(M+1)/N}^{1}(f_0+r_c(iN))} = \frac{c \cdot \gamma}{\alpha(\gamma-1)} \cdot \frac{N(e^{-\frac{\alpha M}{N}}-e^{-\alpha})+(r_0-r_m)\frac{\beta^M(e^{-\frac{\alpha M}{N}}-\beta^{N-M}e^{-\alpha})}{1-\beta}}{(N-M)+(r_0-r_m)\frac{\beta^M(1-\beta^{N-M})}{1-\beta}} \end{cases}$$
(3.9)

**Proposition 1:** In the scenario of reliability learning by WOM, the inequality of the prices in the two stages, $p_1 > p_2$, holds.
**Proof:** The proof is provided in Appendix E.

Proposition 1 shows that the manufacture still makes a markdown pricing decision in the second stage with WOM. That is to say, customers' WOM is not the primary reason for the pricing markdown. By substituting the expression of the optimal price indicated in Equation (3.9) into the expression for the profit indicated in Equation (3.5), we obtain the expressions for the profits in the two sales stages and simplify them as follows:

$$\begin{cases} \pi_1 &= \sum_{j=1}^{M} \frac{r_c(j)+f_0}{Np_1^{\gamma}} p_1 - \frac{r_c(j)+f_0}{Np_1^{\gamma}} ce^{-\frac{\alpha j}{N}} \\ &= \frac{[\alpha(\gamma-1)]^{\gamma-1}}{\gamma^{\gamma} c^{\gamma-1}} \cdot \frac{\left(\frac{M}{N}+\frac{(1-\beta^M)(r_0-r_m)}{N(1-\beta)}\right)^{\gamma}}{\left((1-e^{-\alpha\frac{M}{N}})+\frac{\left(1-\beta^M e^{-\frac{\alpha M}{N}}\right)(r_0-r_m)}{N(1-\beta)}\right)^{\gamma-1}} \\ \pi_2 &= \sum_{j=M+1}^{N} \frac{r_c(j)+f_0}{Np_1^{\gamma}} p_2 - \frac{r_c(j)+f_0}{Np_2^{\gamma}} ce^{-\frac{\alpha j}{N}} \\ &= \frac{[\alpha(\gamma-1)]^{\gamma-1}}{\gamma^{\gamma} c^{\gamma-1}} \cdot \frac{\left(\frac{N-M}{N}+\frac{\beta^M(1-\beta^{N-M})(r_0-r_m)}{N(1-\beta)}\right)^{\gamma}}{\left(\left(e^{-\frac{\alpha M}{N}}-e^{-\alpha}\right)+\frac{\beta^M\left(e^{-\frac{\alpha M}{N}}-\beta^{N-M}e^{-\alpha}\right)(r_0-r_m)}{N(1-\beta)}\right)^{\gamma-1}} \cdot \end{cases}$$
(3.10)

To determine the optimal solution for the total profit of the manufacturer, we first compare total profit ($\pi_1 + \pi_2$) in Equations (3.7) and (3.10). The difference between Equations (3.7) and (3.10) is that (3.10) includes the WOM effect. According to the expression of the WOM in (3.3), the customers' estimation ($r_c$) is a stepwise function over time, and it converges to real reliability. Moreover, within each small sales period, the reliability estimation ($r_c$) is considered to be constant. Therefore, the total number of periods $N$ influences the complexity of solving the optimization problem.

For this $N$-period discrete problem, the number of feasible solutions is $N-1$. If $N$ is not large, the discrete problem can be solved by enumerating every feasible solution from 1 to $N-1$ and comparing the profits to choose the optimal $M^*$ to maximize the total profits. Otherwise, if $N$ is large, it is a complex process to run $N-1$ iterations to calculate the total profits based on $M$ from 1 to $N-1$. Therefore, in Theorem 3 we propose a theoretical approximation method to solve the case when $N$ is large.

**Theorem 3:** When the total number of learning periods $N$ is sufficiently large, we can find a unique and optimal $M \in (0, N)$ to maximize the total profit for the manufacturer.
**Proof:** The proof is provided in Appendix F.

Theorem 3 indicates that when the total number of learning periods ($N$) is large, the solution of the optimization problem with WOM converges to that in the case without WOM. Because $N$ reflects the information diffusion frequency by WOM. A larger $N$ corresponds to a higher frequency of the real reliability information provided by WOM. If the frequency tends to infinity, customers become aware of the reliability promptly.

Consequently, customers know the real reliability, and the case is equivalent to the basic model described in Subsection 3.1.

Furthermore, the total number of learning periods ($N$) also influences the optimality of the problem, because discretizing the feasible region reduces the optimality of the problem in certain cases (Fisher, 1981). Since we discretize the total sales period to describe the WOM, the difference in the continuous and discrete models needs to be analyzed. The special case that customers' initial perceived reliability is equal to the actual reliability of the products is considered as a benchmark because the optimal set of the solutions in the continuous model has been determined.

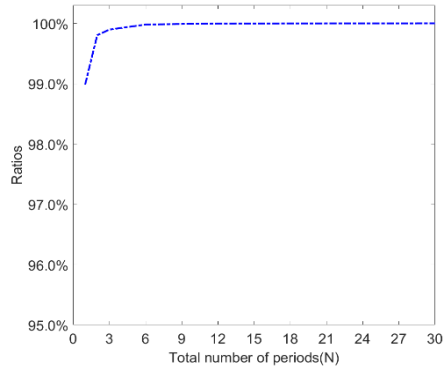

**Figure 3**  Ratio of the total profit in the discretization model to the optimal profit

Figure 3 illustrates the optimal profit for the manufacturer based on the different number of periods the sales process is divided. The vertical coordinate in Figure 3 is the ratio of the total profit in the discrete model to that in the continuous model. When $N$ approaches infinity, the discrete model is equivalent to the continuous model. The optimal total profit increases with $N$ because of the increasing accuracy of the solution. From Figure 3, we find that when $N \geq 6$, the discrete model is almost as good as the continuous model. Furthermore, when $N > 15$, the profit in the discretization model is almost equal to the optimal value. In other words, as $N$ increases, the optimal profit in the discrete model rapidly converges to that in the continuous model. Therefore, we can almost neglect the influence of discretization on the optimality in our model.

**3.4 Effects of WOM**

In this subsection, we discuss the influence of WOM on the manufacturer's total profit. According to theorem 3, when the number of total learning period $N$ approaches infinity, the general case with WOM is equivalent to the special case without WOM. Otherwise, in a finite learning period, by comparing the total profits in Equations (3.7) and (3.10), we clarify the effects of WOM on the total profits in Theorem 4.

**Theorem 4 (Effects of WOM on the total profit):** The customers' learning by WOM can lead to extra profits if and only if $\frac{A_1}{A_1+C_1} + \frac{A_2}{A_2+C_2} < (\frac{A_1+C_1}{B_1+D_1} \cdot \frac{B_1}{A_1})^{\gamma-1} + (\frac{A_2+C_2}{B_2+D_2} \cdot \frac{B_2}{A_2})^{\gamma-1}$, where $A_1 = \frac{M}{N}$, $A_2 = \frac{N-M}{N}$, $B_1 = 1 - e^{-\alpha\frac{M}{N}}$, $B_2 = (e^{-\alpha\frac{M}{N}} - e^{-\alpha})$, $C_1 = \frac{(1-\beta^M)(r_0-r_m)}{N(1-\beta)}$, $C_2 = \frac{\beta^M(1-\beta^{N-M})(r_0-r_m)}{N(1-\beta)}$, $D_1 = \frac{\left(1-\beta^M e^{-\frac{\alpha M}{N}}\right)(r_0-r_m)}{N(1-\beta)}$ and $D_2 = \frac{\beta^M\left(e^{-\frac{\alpha M}{N}} - \beta^{N-M}e^{-\alpha}\right)(r_0-r_m)}{N(1-\beta)}$.

**Proof:** The proof is provided in Appendix G.

Theorem 4 specifies the sufficient and necessary conditions to compare the profits in the two cases. If we define the basic case without WOM as the benchmark, we find that the underestimation (UE) of the reliability always leads to lower profits. In contrast, when customers overestimate the reliability, the manufacturer tends to attain higher profits than the benchmark. However, the overestimation (OE) does not always lead to extra profits. Because the profit depends not only on customers' initial estimation but also on the number of total learning periods ($N$). To ensure that OE can achieve additional benefits for the manufacturer, Proposition 2 provides a sufficient condition for the OE to achieve higher profits than the benchmark.

**Proposition 2:** If the customers' initial perception is higher than the real reliability, one of the sufficient conditions of WOM to achieve higher profits is $\frac{A_1}{B_1} < \frac{C_1}{D_1}$ $\left(\frac{1-\beta^M e^{-\frac{\alpha M}{N}}}{1-\beta^M} < \frac{1-e^{-\alpha\frac{M}{N}}}{\frac{M}{N}}\right)$ and $\frac{A_2}{B_2} < \frac{C_2}{D_2}$ $\left(\frac{e^{-\frac{\alpha M}{N}} - \beta^{N-M}e^{-\alpha}}{1-\beta^{N-M}} < \frac{e^{-\frac{\alpha M}{N}} - e^{-\alpha}}{\frac{N-M}{N}}\right)$.

**Proof:** The proof is provided in Appendix H.

In the overestimation scenario, larger $\beta$ or smaller $N$ more likely satisfy the conditions in Proposition 2. Larger $\beta$ means that customers learn a smaller amount of real information each time, and smaller $N$ corresponds to smaller learning frequency. Under either of these conditions, customers obtain a smaller amount of information from WOM. In other words, letting customers know the real reliability can gradually lead to additional profits. Furthermore, the difference in the initial perception of reliability ($r_0$) and the real reliability ($r_m$) considerably influences the total profit. A larger difference ($r_0 - r_m$) corresponds to a higher profit attained by the manufacturer (see Section 4.4 for a detailed discussion).

Figures 4 and 5 show the change in the total profit based on different learning depths $(1-\beta)$ and the number of learning periods $(N)$ in the overestimation and underestimation conditions, respectively. Furthermore, the comparison of these two cases with the benchmark is also shown in these two figures.

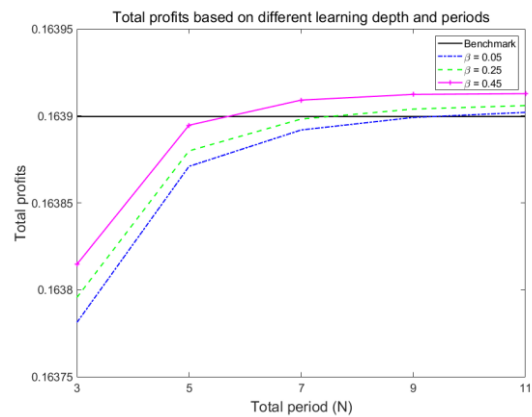

**Figure 4**　Overestimation (OE) scenario

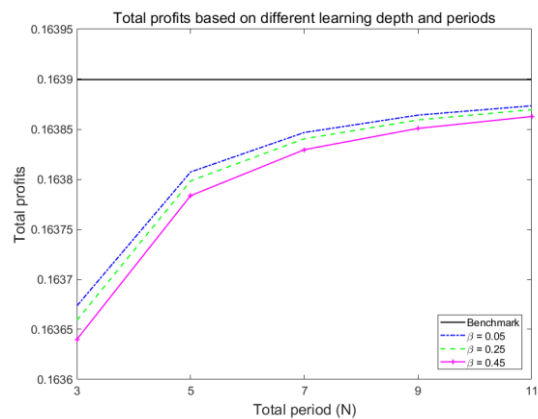

**Figure 5**　Underestimation (UE) scenario

Figure 4 illustrates the case when the customers' perception of product reliability is higher than the real reliability. In this case, if the total number of learning periods is small, which corresponds to fewer opportunities to switch prices, the overestimation may not cover the loss of the profits owing to the limited choice, and the total profit is less than those in the benchmark scenario. Furthermore, if customers obtain a larger amount of real information in each period (smaller $\beta$), the rate of convergence to the real reliability is higher, and the manufacturer achieves smaller extra profits.

Figure 5 corresponds to the case when the customers' initial perception reliability is lower than the real reliability. In this circumstance, regardless of the changes in the learning process, the total profit decreases owing to the effects of the negative WOM. In contrast, for the overestimation scenario, if customers obtain a larger amount of real information in each period ($\beta$ is smaller) and learn more frequently ($N$ is larger), the manufacturer loses fewer profits. In this case, the manufacturer must let the real reliability be known as soon as possible to reduce profits loss.

## 4. Extended model with warranty

In this section, we introduce warranty as another factor that may affect WOM and investigate the influence of their interactions on the selling prices, switch point of the two stages, and total profit.

Warranty is usually an obligation of manufacturers to repair or replace a product when it fails during the promissory period. A basic warranty is often included in the sales prices, but customers must pay for an extended warranty additionally. Therefore, we focus on the extended warranty during the lifetime of products. The price of the extended warranty is denoted as $p_w$, which is also decided by the manufacturer. Customers have the choice of spending $p_w$ to buy the warranty or not buying.

From the perspective of customers, the warranty price not only affects the demand for a warranty but also indirectly reflects the reliability of products. Therefore, the warranty price exhibits interaction with WOM as well.

### 4.1 Influence of the warranty service

Similar to that in Section 3, WOM is reflected as customers' perception of reliability. Considering the influence of the warranty, the customers' perceived reliability is

$$r_c(i) = \begin{cases} \beta_0 r_0 + (1 - \beta_0) \cdot (a - bp_w) & \text{for } i = 0 \\ \beta_1 r_c(i-1) + (1 - \beta_1) r_m & \text{for } i \geq 1, \end{cases}$$

(4.1)

where $\beta_0$ and $\beta_1$ denote the smooth factors of the initial period and other periods, respectively, and we have $\beta_0, \beta_1 \in (0,1)$ based on the works of Campbell (2015) and Lei et al. (2017), Moreover, $b > 0$ is the sensitivity parameter of the customer's initial perceptions of reliability over the warranty price. According to Crosby (1980), if a manufacturer is confident in the reliability of his/her products, the warranty can be presented for free. Therefore, a boundary condition is specified in Remark 1, according to practical considerations.

**Remark 1:** *When $p_w = 0$, we assume that from the perspective of customers, the products are reliable, because the manufacturer is confident in the quality of his/her products and willing to repair them freely during the product lifetime. In this case, customers can estimate the reliability as 1. Thus, $\beta_0 r_0 + (1 - \beta_0) \cdot a = 1$, and $a = \frac{1 - \beta_0 r_0}{1 - \beta_0}$.*

According to Li (2011), the probability of customers purchasing warranty is defined as

$$P = 1 - d \cdot p_w, \quad (4.2)$$

where $d$ represents the sensitivity of the warranty price in terms of the warranty demand. We assume that the choice to buy the extended warranty service is independent of the usage behavior of customers and the failure of the products. Based on this assumption, the total number of failure claims during the two stages is expressed as

$$N_{Tot\_claim} = P \cdot f_0 \cdot \left( \frac{\int_0^\theta \lambda(t)(1 + r_c(t) - r_m)dt}{p_1^\gamma} + \frac{\int_\theta^1 \lambda(t)(1 + r_c(t) - r_m)dt}{p_2^\gamma} \right). \quad (4.3)$$

With the extended warranty, the manufacturer's total profit consists of the sales profit and warranty profit. According to the demand for products in Equation (3.2) and the probability of purchasing the warranty in Equation (4.2), we have the total profit in two stages as follows:

$$\begin{cases} \pi_1(p_1, \theta, p_w) = \frac{\sum_{i=1}^{M} \int_{i-1}^{i} \lambda(t)(1 + r_c(i) - r_m)(p_1 - c \cdot e^{-\alpha t})dt}{p_1^\gamma} + \frac{\sum_{i=1}^{M} \int_{i-1}^{i} \lambda(t)(1 + r_c(i) - r_m)(1 - dp_w)(p_w - c_w f_0)dt}{p_1^\gamma} \\ \pi_2(p_2, \theta, p_w) = \frac{\sum_{i=M+1}^{N} \int_{i-1}^{i} \lambda(t)(1 + r_c(i) - r_m)(p_2 - c \cdot e^{-\alpha t})dt}{p_2^\gamma} + \frac{\sum_{i=M+1}^{N} \int_{i-1}^{i} \lambda(t)(1 + r_c(i) - r_m)(1 - dp_w)(p_w - c_w f_0)dt}{p_2^\gamma} \end{cases}. \quad (4.4)$$

Using the same method as that in Section 3, the product prices in two stages are expressed as

$$\begin{cases} p_1(M, p_w) = \frac{\gamma \cdot \left[ c \cdot \sum_{i=1}^{M} e^{-\frac{\alpha i}{N}}(r_i + f_0) - (1 - dp_w) \cdot (p_w - f_0 c_w) \sum_{i=1}^{M}(f_0 + r_i) \right]}{(\gamma - 1) \sum_{i=1}^{M}(f_0 + r_i)} \\ p_2(M, p_w) = \frac{\gamma \cdot \left[ c \cdot \sum_{i=M+1}^{N} e^{-\frac{\alpha i}{N}}(r_i + f_0) - (1 - dp_w) \cdot (p_w - f_0 c_w) \sum_{i=M+1}^{N}(f_0 + r_i) \right]}{(\gamma - 1) \sum_{i=M+1}^{N}(f_0 + r_i)} \end{cases}. \quad (4.5)$$

Owing to the complexity of this extended model, it is difficult to prove the optimality and uniqueness theoretically. Thus, we numerically demonstrate the optimality.

## 4.2 Numerical illustration

To discuss the patterns of the key variables in the extended model, we conduct a numerical analysis. Table 2 illustrates the values of non-decision parameters, based on actual conditions. For example, according to Kim and Park (2008), the warranty cost for products is 20% of the production cost on average, and the average failure rate of consumer electronics is approximately 10%. Thus, the warranty cost for the product ($c_w$) is set as 0.2, the average failure rate ($f_0$) is set as 0.1, and the production cost ($c$) is normalized to 1.

**Table 2**  Numerical values of the parameters in the extended model

| Parameter | $\alpha$ | $\beta_0$ | $\beta_1$ | $\gamma$ | $f_0$ | $c$ | $c_w$ | $d$ | $b$ | $N$ | $r_0$ |
|---|---|---|---|---|---|---|---|---|---|---|---|
| Value | 0.1 | 0.2 | 0.5 | 3 | 0.1 | 1 | 0.2 | 5 | 5 | 200 | 0.8 |

Based on the values in Table 2, we can determine the optimal warranty price $p_w^* = 0.1092$. In this setting, we find a unique optimal solution to maximize the total profits. Furthermore, the optimal switching time for price change is $\theta^* = \frac{102}{200}$, and the optimal prices in the two stages are $p_1 = 1.4012$ and $p_2 = 1.3300$. Figure 6 illustrates the marginal effects of the warranty price on total profits. Each point on the curve denotes the maximum profits by optimizing $p_1$, $p_2$ and $\theta$ based on a given warranty price. Figure 6 indicates that there exists a unique warranty price for the manufacturer to jointly optimize the total profits.

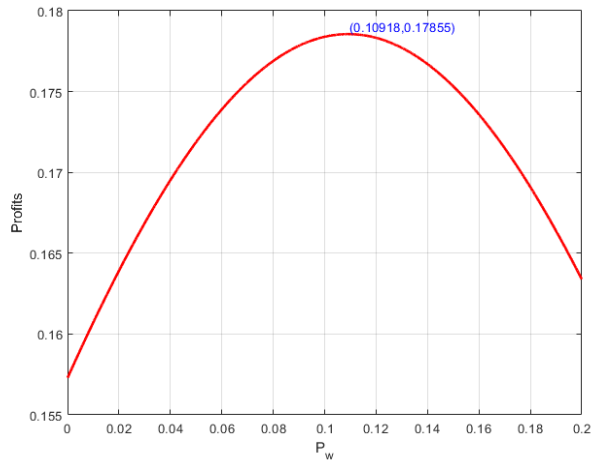

**Figure 6**  Relationship of the total profits and warranty price

## 4.3 Comparison of different cases based on WOM and warranty

In this subsection, we compare four cases through numerical experiments: Case I does not include WOM or warranty; Case II and III only take into account the WOM and the warranty, respectively; and Case IV involves both the WOM and warranty. To distinguish positive and negative WOM, we consider both OE and UE conditions in Cases II and IV. Table 3 presents product prices in the two stages, price switching time, warranty price, and total profits in the four cases.

Table 3　Numerical results for the four cases

| Case | | Price | | Price switching time $\theta^*$ | Warranty price $p_w$ | Profits (Percentage of Case I) |
|---|---|---|---|---|---|---|
| | | $p_1$ | $p_2$ | | | |
| I: Without WOM and warranty | | 1.4625 | 1.3912 | 0.50833 | — | 0.1639 (100%) |
| II: With WOM only | OE | 1.4590 | 1.3876 | 13/25 | — | 0.1645 (100.4%) |
| | UE | 1.4585 | 1.3876 | 13/25 | — | 0.1633 (99.6%) |
| III: With warranty only | | 1.4017 | 1.3304 | 0.50908 | 0.1100 | 0.1788 (109.1%) |
| IV: With both WOM and warranty | OE | 1.3976 | 1.3272 | 13/25 | 0.1033 | 0.1767 (107.8%) |
| | UE | 1.3976 | 1.3272 | 13/25 | 0.1033 | 0.1767 (107.8%) |

　　Table 3 indicates that the optimal warranty strategy increases the total profit of the manufacturer in both the positive and negative WOM cases. This result demonstrates the profitability of the warranty, and the total benefit increases by approximately 10%. Comparing with a warranty, WOM exerts only a slight influence on the sales process and total profits. In terms of the impact of warranty on WOM, because the warranty price reflects the reliability information of products, the difference between OE and UE cases is small. Consequently, the optimal prices and profits are similar in these two cases. Finally, in terms of the selling price of products, a warranty considerably reduces the prices. Because warranty can help achieve additional profits, the manufacturer reduces the products' price and increase the sales volume, which increases the warranty sales.

**Observation 3:** *The decreasing trend of prices over the learning effect of the manufacturer (α) in Case II (the case with WOM only) is similar to Case IV (the case* *with both WOM and warranty). Meanwhile, the markdown value $\Delta p = p_1 - p_2$ increases with α, and the prices in Case IV are lower than those in Case II.*

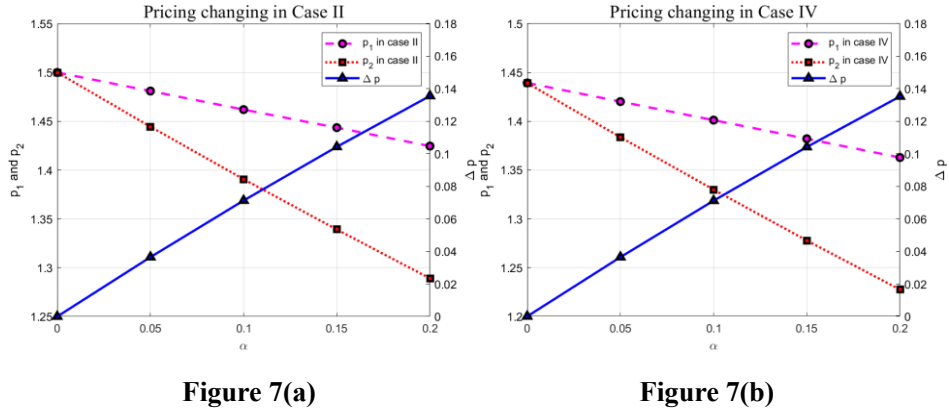

**Figure 7(a)**          **Figure 7(b)**

**Figure 7**  Changes of optimal prices in two stages and their difference with learning effect α

Figures 7 (a) and (b) indicate that the decrease trends in the two cases are similar. The prices decrease with the learning effect parameter α because a larger α means the manufacturing cost decreases more rapidly over time. Furthermore, the markdown value ($\Delta p$) increases with the learning effect parameter α. When the manufacturing cost reduces rapidly, the manufacturer can lower the price to stimulate consumption and achieve higher demand and profits. By comparing Figure 7(a) with 7(b), it can be noted that the price level in Case IV is lower than that in Case II. Because the warranty can achieve additional profits for the manufacturer, and a lower price can lead to an increased demand for the products and warranty.

**Observation 4:** *The decreasing trend of price over the price elasticity parameter (γ) is exponential both in Case II and Case IV. Similar to learning effect parameter α, the prices in Case IV are lower than those in Case II. However, different from α, the markdown value also decreases exponentially.*

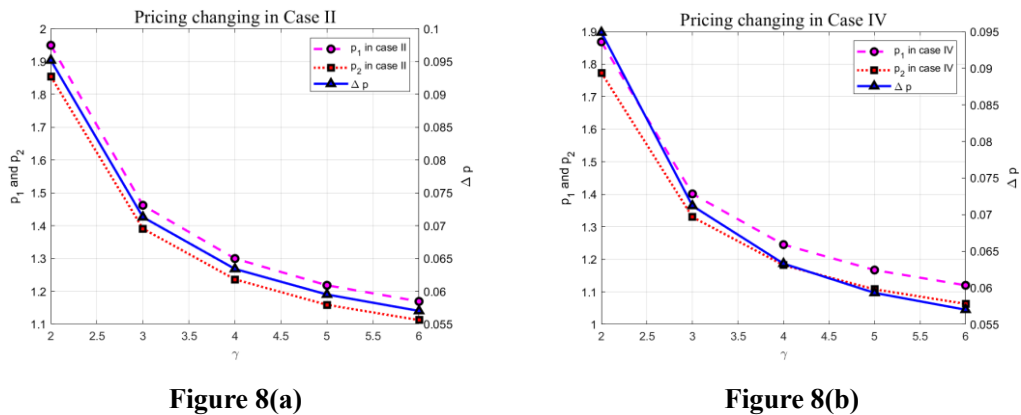

**Figure 8(a)**          **Figure 8(b)**

**Figure 8**  Changes of optimal prices in two stages and their difference with price elasticity γ

Figures 8(a) and 8(b) illustrate the influence of price elasticity parameter $\gamma$ in Case II and Case IV, respectively. For larger price elasticity, as the product prices increase, the sales volumes decrease more sharply. Similar to the influence of learning effect parameter $\alpha$, the price level in Case IV is lower than that in Case II. As shown in Figures 8(a) and 8(b), the optimal $p_1$ and $p_2$ decrease exponentially with $\gamma$. Moreover, the markdown value $\Delta p = p_1 - p_2$ also decreases similarly.

**Observation 5:** *Since internal (i.e. the learning effect of the manufacturer) and external factors (i.e. customers' price elasticity) can result in the decrease of prices, only the internal factors lead to a voluntary pricing markdown of the manufacturer, and increase the total profits. Moreover, the warranty can relieve the effect of customers' price elasticity over profit.*

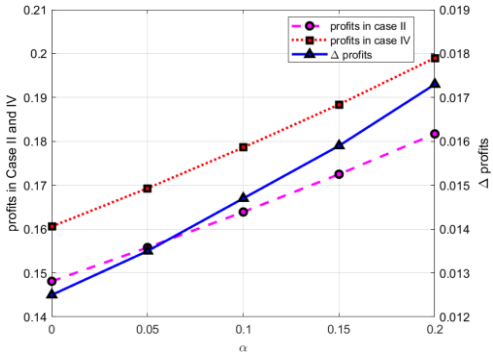
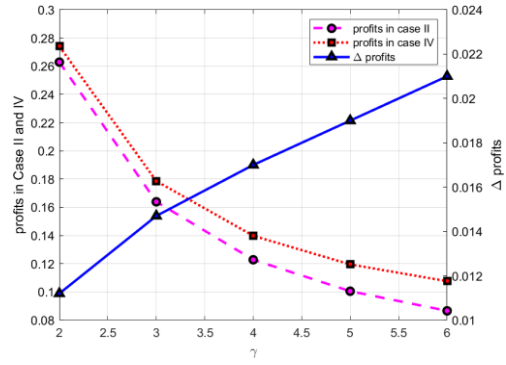

**Figure 9(a)**  **Figure 9(b)**

**Figure 9**  Profit changes in the two cases and their difference with learning effect parameter $\alpha$ and price elasticity parameter $\gamma$

Figure 9(a) illustrates that stronger learning effects and optimal warranty service lead to higher profits. Moreover, warranty strategy leads to more significant advantages for the manufacturer under higher learning abilities, because the difference in the profits of Cases II and IV increases with learning effect parameter $\alpha$. Figure 9(b) illustrates the influence of the elasticity between the demand and price of customers. Given a certain price, a larger price elasticity parameter $\gamma$ corresponds to a smaller sales volume. Therefore, the total profit decreases with $\gamma$. We also find that total profit in Case IV is always higher and the difference between Cases I and IV increases with $\gamma$. In other words, the warranty plays an important role when customers are sensitive toward the prices because the manufacturer can lower the prices to increase the sales volume and attain extra profits through the warranty sales.

## 4.4 Interaction effects of WOM and warranty

As described in Section 3, the difference between the real reliability and perceived reliability of customers leads to a fluctuation in sales volume. Because in practice, the real reliability is usually a latent factor that is unknown to neither the manufacturer nor customers. In our setting, the real reliability is a constant hidden factor. Therefore, based on the linear smoothing method for the WOM defined in Equation (3.3), Figure 10 shows the convergence path of the perceived reliability by customers ($r_c$). It can be observed that (a) the difference in the real reliability and $r_c$ depends considerably on the initial perceptions of reliability, and (b) each path converges monotonically to the real reliability.

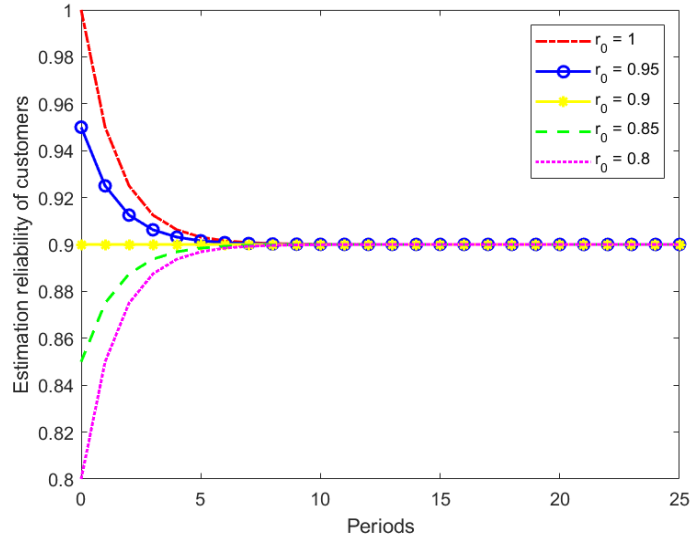

**Figure 10** Convergence for different initial reliabilities

Based on the different convergence paths, the influence of different initial perceptions of reliability on the profits and optimal warranty price is illustrated in Figure 11. The increased trends of the profits in Case II and IV are similar because a higher reliability perception increases the sales volume. Interestingly, if customers initially appraise the reliability of a product as high, the warranty deserves a high value as well. This phenomenon can explain the feedback mechanism of the warranty price and reliability. Given a certain initial reliability perception of customers ($r_0$), the warranty price negatively influences the reliability perception of customers ($r_c$). However, the initial perception reliability perception ($r_0$) positively influences the warranty price, and $r_0$ is also a key aspect related to the reliability perception of customers ($r_c$). This feedback mechanism ensures the stability of the warranty price.

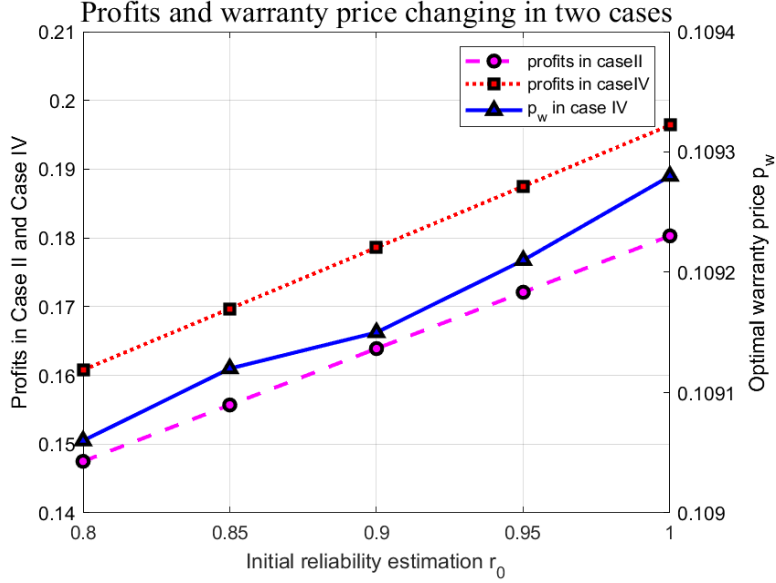

**Figure 11**  Change in the profits and optimal warranty price with $r_0$

## 5. Conclusions

In this paper, we focus on the problem for a monopolist manufacturer to seek the optimal pricing strategy for nondurable products, especially consumer electronics. We propose a model considering both internal factors (i.e. learning effect) and external factors (i.e. WOM, customers' price elasticity) to maximize the manufacturer's total profit. As a result, we theoretically optimize the price in the two sales and the price switching time and prove the uniqueness and optimality of the solutions.

For internal factors, the learning effect is important for manufacturers to reduces the manufacturing cost. We find that the larger the learning parameter is, the more the selling price decreases, and the more profits the manufacturer gets. In contrast, for external factors, the price elasticity parameter reflects the customers' attitude toward the price. If customers are sensitive toward the price, the manufacturer will set a lower price to guarantee sales volume, but the total benefit still decreases.

Another key factor for consumer electronics is the warranty. Warranty is an important value-added service for the sales of electronics. Moreover, the warranty also has interactions with WOM because the warranty price can reflect the actual reliability and influence the WOM. Based on the extended model considering both the warranty and WOM, we jointly optimize the prices of the products in the two stages, price switching time and warranty price. By comparing the models considering different factors in four cases, we conclude that the optimal warranty service leads to additional profits (approximately 10%) for the manufacturer. Moreover, from the perspective of customers, the warranty price is an indicator of the product's reliability. Consequently, the difference (including the prices in the two stages, price change time, and total profits) between overestimation and underestimation cases shrinks in the case with warranty service.

Based on theoretical results, we provide managerial insights as follows. First, for WOM, by setting the scene in which customers' initial perceived reliability equals actual reliability as a benchmark, we find the customer's overestimation does not always lead to extra profits while the underestimation by the customer always leads to reduced profits. Thus, the manufacturer needs to pay more attention to striving and publicity their positive WOM for their products. Second, for the warranty, it is not only one of the most important channels for the manufacturer to improve its profits, but also a signal for customers to evaluate the actual reliability of products. Moreover, the feedback mechanism of the warranty price and the customers' perceived reliability exists, which leads to the stability of the warranty price. Finally, for the price of products and total profit, although customers' price elasticity and the manufacturer's learning effects both reduce the price, only the internal learning leads to initiative pricing markdown and more profits. Since the external factors are hard to control, the manufacture would pay more attention to improve its technology and lower the manufacturing cost.

Our work can be extended in several future directions. For example, one can generalize the two-stage pricing policy as a multistage pricing policy. Second, the different types of failure rates can be considered to analyze the case when the customers' perceived reliability does not converge to the real reliability. Finally, in our work, customers were treated as a group. The differences between individual customers can be considered in future research.

# Appendix

**Appendix A. Proof of Lemmas:**

In order to prove the optimality of the solution in theorem 1, we present several lemmas to be used later.

**Definition 1:** Define

$$f(x) = e^{-\alpha x}. \qquad (A.1)$$

And we have $f'(x) = -\alpha e^{-\alpha x}$ and $f''(x) = \alpha^2 e^{-\alpha x}$.

*The absolute slope values of the secant for* $f(x)$ *are defined as* $\varphi_1(x) = \frac{(1-e^{-\alpha x})}{x}$ *and*

$\varphi_2(x) = \frac{(e^{-\alpha x} - e^{-\alpha})}{1-x}$. *The difference in the $\gamma^{th}$ power of these values is $\varphi_\Delta(x, \gamma) = \varphi_1^\gamma - \varphi_2^\gamma$. Geometrically, $\varphi_1$ and $\varphi_2$ denote the absolute slope values of the lines joining two points $(0, f(0))$ and $(x, f(x))$ and $(x, f(x))$ and $(1, f(1))$, respectively.*

**Lemma 1.** $\varphi_1(x) > \varphi_2(x) > 0$.
**Proof:** The proof is provided in Appendix A.1.

**Lemma 2.** $\varphi_1'(x) < 0 \quad \varphi_2'(x) < 0$
**Proof:** The proof is provided in Appendix A.2.

**Lemma 3.** *The upper bound of $\varphi_2(x)$ is equal to the lower bound of $\varphi_1(x)$, and $\min_{x \in (0,1)} \varphi_1(x) = \max_{x \in (0,1)} \varphi_2(x) = (1 - e^{-\alpha})$. The bounds cannot be achieved simultaneously.*
**Proof:** The proof is provided in Appendix A.3.

**Lemma 4.** *Based on the definition of $\varphi_1(x)$, $\varphi_2(x)$ and $f(x)$ for $x \in (0,1)$, we have $\varphi_1(x) > \left|f'\left(\frac{x}{2}\right)\right|$, and $\varphi_2(x) > \left|f'\left(\frac{1+x}{2}\right)\right|$.*
**Proof:** The proof is provided in Appendix A.4.

**Lemma 5.** *When $\gamma \geq 2$, $\varphi_\Delta(x, \gamma)$ is decreasing in x.*
**Proof:** The proof is provided in Appendix A.5.

**Definition 2:** Let $z(x) = (\gamma - 1)\alpha e^{-\alpha x}(\varphi_1^\gamma - \varphi_2^\gamma) - \gamma(\varphi_2 \varphi_1^\gamma - \varphi_1 \varphi_2^\gamma)$, and
$\begin{cases} z_1(x, \gamma) = \alpha e^{-\alpha x}(\varphi_1^\gamma - \varphi_2^\gamma) \\ z_2(x, \gamma) = \varphi_2 \varphi_1^\gamma - \varphi_1 \varphi_2^\gamma \end{cases}$;
*consequently, $z(x) = (\gamma - 1)z_1(x, \gamma) - \gamma z_2(x, \gamma)$.*

**Lemma 6.** *Define* $k(x) = \frac{\alpha(1-e^{-\alpha x})}{2x} - \frac{\frac{1-e^{-\alpha x}}{x} - \alpha e^{-\alpha x}}{x} - \frac{\alpha(e^{-\alpha x} - e^{-\alpha})}{2(1-x)} + \frac{\alpha e^{-\alpha x} - \frac{e^{-\alpha x} - e^{-\alpha}}{1-x}}{1-x}$,

*where* $k_1(x) = \frac{\alpha(1-e^{-\alpha x})}{2x} - \frac{\frac{1-e^{-\alpha x}}{x} - \alpha e^{-\alpha x}}{x}$ *and* $k_2(x) = -\frac{\alpha(e^{-\alpha x} - e^{-\alpha})}{2(1-x)} +$

$\frac{\alpha e^{-\alpha x} - \frac{e^{-\alpha x} - e^{-\alpha}}{1-x}}{1-x}$. *Consequently,* $\begin{cases} k(x) = k_1(x) + k_2(x) > 0 \\ k_1(x) > 0 \end{cases}$.

**Proof:** The proof is provided in Appendix A.6.

**Lemma 7 (Monotonicity of the function $z(x)$).** $z(x)$ *decreases in x when* $x \in (0,1)$.

**Proof:** The proof is provided in Appendix A.7.

**Lemma 8 (Boundary conditions of $z(x)$).** $z(x) = 0$ *has the unique solution* $x^* \in (0,1)$. *When* $x \in (0, x^*)$, $z(x) > 0$; *otherwise,* $x \in (x^*, 1)$, $z(x) < 0$.

**Proof:** The proof is provided in Appendix A.8.

### A.1. Proof of Lemma 1

Based on Definition 1, we have the formulation:
$$\begin{cases} \varphi_1(x) = \frac{1-e^{-\alpha x}}{x} \\ \varphi_2(x) = \frac{e^{-\alpha x}-e^{-\alpha}}{1-x}. \end{cases} \quad (A.1.1)$$

Because the range of $x$ is $x \in (0,1)$, we have that $1 - e^{-\alpha x} > 0$ and $e^{-\alpha x} - e^{-\alpha} > 0$. Thus $\varphi_1(x) > 0$, $\varphi_2(x) > 0$.

According to the Lagrange mean value theorem and the definition of $f(x)$, there exist $\xi \in (0, x)$, and $\eta \in (x, 1)$, such that:
$$\begin{cases} f'(\xi) = \frac{f(x)-f(0)}{x-0} = -\frac{1-e^{-\alpha x}}{x} = -\varphi_1(x) \\ f'(\eta) = \frac{f(1)-f(x)}{1-x} = -\frac{e^{-\alpha x}-e^{-\alpha}}{1-x} = -\varphi_2(x) \end{cases} \quad (A.1.2)$$

According to Equation (A.1), we have $f''(x) = \alpha^2 e^{-\alpha x} > 0$. Because $\xi < \eta$, we have $f'(\xi) < f'(\eta)$. Thus, we have $\varphi_1(x) > \varphi_2(x)$ based on (A.1.2).

### A.2. Proof of Lemma 2

Based on Equation (A.1.2), we have the first order derivatives of $\varphi_1(x)$ and $\varphi_2(x)$ are:
$$\begin{cases} \varphi_1'(x) = -\frac{\frac{1-e^{-\alpha x}}{x}-\alpha e^{-\alpha x}}{x} = -\frac{f'(x)-f'(\xi)}{x} \\ \varphi_2'(x) = -\frac{\alpha e^{-\alpha x}-\frac{e^{-\alpha x}-e^{-\alpha}}{1-x}}{1-x} = \frac{f'(x)-f'(\eta)}{1-x}. \end{cases} \quad (A.2.1)$$

Equation (A.1) shows that $f''(x) > 0$. Because $\xi \in (0, x)$, $\eta \in (x, 1)$, we have
$$f'(\xi) < f'(x) < f'(\eta). \quad (A.2.2)$$

Hence, we have $\varphi_1'(x) < 0$, and $\varphi_2'(x) < 0$.

## A.3. Proof of Lemma 3

Based on Lemma 2, we know that $\varphi_1(x)$ and $\varphi_2(x)$ are decreasing functions. Thus we have $\min_{x \in (0,1)} \varphi_1(x) = \varphi_1(1) = 1 - e^{-\alpha}$. Similarly, we have $\max_{x \in (0,1)} \varphi_2(x) = \varphi_2(0) = (1 - e^{-\alpha})$. And the bounds of these two functions cannot be reached simultaneously.

## A.4. Proof of Lemma 4

To begin with, based on (A.1), we have

$$\begin{cases} \left|f'\left(\frac{x}{2}\right)\right| = \alpha e^{-\frac{\alpha x}{2}} \\ \left|f'\left(\frac{1+x}{2}\right)\right| = \alpha e^{-\frac{\alpha(1+x)}{2}}. \end{cases} \quad (A.4.1)$$

(1) Firstly, we show that $\varphi_1(x) > \left|f'\left(\frac{x}{2}\right)\right|$.

Define

$$g_1(x) = (1 - e^{-\alpha x}) - \alpha x e^{-\frac{\alpha x}{2}}. \quad (A.4.2)$$

It is easy to find that $g_1(0) = 0$. The first order derivative of $g_1(x)$ is:

$$g_1'(x) = \alpha e^{-\alpha x} - \alpha e^{-\frac{\alpha x}{2}} + \frac{\alpha^2 x}{2} e^{-\frac{\alpha x}{2}} = e^{-\frac{\alpha x}{2}} g_2(x), \quad (A.4.3)$$

where

$$g_2(x) := \alpha e^{-\frac{\alpha x}{2}} - \alpha + \frac{\alpha^2 x}{2}. \quad (A.4.4)$$

We have $g_2(0) = 0$, and the derivative of $g_2(x)$ is:

$$g_2'(x) = -\frac{\alpha^2}{2} e^{-\frac{\alpha x}{2}} + \frac{\alpha^2}{2} = \frac{\alpha^2}{2}\left(1 - e^{-\frac{\alpha x}{2}}\right) > 0.$$

Thus, when $x \in (0,1)$, we have $g_2(x) > g_2(0) = 0$. Hence, based on (A.4.3) and (A.4.4), we have $g_1'(x) > 0$. As a result, we have $g_1(x) > g_1(0) = 0$.

Furthermore, based on (A.4.2), we have $(1 - e^{-\alpha x}) > \alpha x e^{-\frac{\alpha x}{2}}$.

Finally, because $x > 0$, we have

$$\varphi_1(x) = \frac{1-e^{-\alpha x}}{x} > \alpha e^{-\frac{\alpha x}{2}} = \left|f'\left(\frac{x}{2}\right)\right|, \quad (A.4.5)$$

where the first equality holds because of Definition 2 and the last equality holds because of (A.4.1).

(2) Secondly, we show that $\varphi_2(x) > \left|f'\left(\frac{1+x}{2}\right)\right|$

Define $h_1(x) = (e^{-\alpha x} - e^{-\alpha}) - (1-x)\alpha e^{-\frac{\alpha(1+x)}{2}}$. It is easy to find that $h_1(1) = 0$.
The first derivative of $h_1(x)$ is:

$$h_1'(x) = -\alpha e^{-\alpha x} + \alpha e^{-\frac{\alpha(1+x)}{2}} + (1-x)\frac{\alpha^2}{2}e^{-\frac{\alpha(1+x)}{2}}$$

$$= e^{-\frac{\alpha x}{2}}\left(-\alpha e^{-\frac{\alpha x}{2}} + \alpha e^{-\frac{\alpha}{2}} + (1-x)\frac{\alpha^2}{2}e^{-\frac{\alpha}{2}}\right)$$

(A.4.6)

The sign of $h_1'(x)$ depends on $-\alpha e^{-\frac{\alpha x}{2}} + \alpha e^{-\frac{\alpha}{2}} + (1-x)\frac{\alpha^2}{2}e^{-\frac{\alpha}{2}}$, which is defined as $h_2(x)$. The first derivative of $h_2(x)$ is:

$$h_2'(x) = \frac{\alpha^2}{2}e^{-\frac{\alpha x}{2}} - \frac{\alpha^2}{2}e^{-\frac{\alpha}{2}} > 0. \tag{A.4.7}$$

Thus, we have that $h_2(x)$ is an increasing function, and so $h_2(x) < h_2(1) = 0$.
Hence, we have $h_1'(x) < 0$ and $h_1(x) > h_1(1) = 0$ for $0 < x < 1$. As a result,

$$(e^{-\alpha x} - e^{-\alpha}) > (1-x)\alpha e^{-\frac{\alpha(1+x)}{2}}. \tag{A.4.8}$$

Because $1 - x > 0$, we have $\varphi_2 = \frac{e^{-\alpha x} - e^{-\alpha}}{1-x} > -f'\left(\frac{1+x}{2}\right) = \left|f'\left(\frac{1+x}{2}\right)\right|$.

## A.5. Proof of Lemma 5

Based on Definition 1, $\varphi_\Delta(x,\gamma) = \varphi_1^\gamma(x) - \varphi_2^\gamma(x)$. The partial derivative of $\varphi_\Delta(x,\gamma)$ over x is:

$$\frac{\partial \varphi_\Delta(x,\gamma)}{\partial x} = \gamma\varphi_1^{\gamma-1}(x)\varphi_1'(x) - \gamma\varphi_2^{\gamma-1}(x)\varphi_2'(x) \tag{A.5.1}$$

According to Lemma 1, we have $\varphi_1 > \varphi_2 > 0$ and so $\gamma\varphi_1^{\gamma-1} > \gamma\varphi_2^{\gamma-1}$ for $\gamma > 1$. Meanwhile, according to Lemma 2, we have $\varphi_1'(x) < 0$ and $\varphi_2'(x) < 0$. Thus the sufficient condition to prove Lemma 5 is $\varphi_1'(x) < \varphi_2'(x)$. Based on (A.2.1), the expression of $\varphi_1'(x) - \varphi_2'(x)$ is

$$\varphi_1'(x) - \varphi_2'(x) = \frac{f'(\xi) - f'(x)}{x} - \frac{f'(x) - f'(\eta)}{1-x}$$

$$= \frac{1}{2}\left(\frac{f'(\xi) - f'(x)}{\frac{x}{2}} - \frac{f'(x) - f'(\eta)}{\frac{1-x}{2}}\right). \tag{A.5.2}$$

According to (A.1), (A.1.2), (A.4.5), and (A.4.8), we have:

$$\begin{cases} -f'(\xi) = \dfrac{1-e^{-\alpha x}}{x} > \alpha e^{-\tfrac{\alpha x}{2}} = -f'\left(\dfrac{x}{2}\right) \\ -f'(\eta) = \dfrac{e^{-\alpha x}-e^{-\alpha}}{1-x} > \alpha e^{-\tfrac{\alpha(1+x)}{2}} = -f'\left(\dfrac{1+x}{2}\right). \end{cases}$$

Because $f''(x) > 0$ due to (A.1), we have $\xi < \tfrac{x}{2}$, $\eta < \tfrac{1+x}{2}$. Hence, we have

$$|\xi - x| > \tfrac{x}{2} \text{ and } |\eta - x| = \eta - x < \tfrac{1-x}{2}. \tag{A.5.3}$$

Thus, based on Equation (A.5.2), we have

$$\varphi_1'(x) - \varphi_2'(x) = \frac{1}{2}\left(\frac{f'(\xi) - f'(x)}{\tfrac{x}{2}} - \frac{f'(x) - f'(\eta)}{\tfrac{1-x}{2}}\right)$$

$$= \frac{1}{2}\left(\frac{f'(\eta) - f'(x)}{\tfrac{1-x}{2}} - \frac{f'(x) - f'(\xi)}{\tfrac{x}{2}}\right)$$

$$< \frac{1}{2}\left(\frac{f'(\eta) - f'(x)}{\eta - x} - \frac{f'(x) - f'(\xi)}{x - \xi}\right) = \frac{1}{2}(f''(\eta_1) - f''(\xi_1)),$$

where $\eta_1 \in [x, \eta]$ and $\xi_1 \in [\xi, x]$, and the first inequality holds because of (A.5.3). Based on (A.1), we have $f'''(x) = -\alpha^3 e^{-\alpha x} < 0$. Thus, $f''(\eta_1) < f''(\xi_1)$ because $\eta_1 \geq x \geq \xi_1$. And so $\varphi_1'(x) - \varphi_2'(x) < 0$.

**A.6. Proof of Lemma 6**

Based on the definition, the first derivate of $k(x)$ is

$$k'(x) = k_1'(x) + k_2'(x) = \frac{(-\alpha^2 x^2 - 3\alpha x - 4)e^{-\alpha x} - \alpha x + 4}{2x^3} + \frac{(4 + \alpha^2(1-x)^2 - 3\alpha(1-x))e^{-\alpha x} - e^{-\alpha}(4 + \alpha(1-x))}{-2(1-x)^3}$$

(A.6.1)

(1) Firstly, we analyze $k_1(x)$.

The boundary condition is:

$$\begin{aligned} \lim_{x\to 0} k_1(x) &= \lim_{x\to 0} \frac{\alpha(1-e^{-\alpha x})}{2x} - \frac{\tfrac{1-e^{-\alpha x}}{x} - \alpha e^{-\alpha x}}{x} \\ &= \lim_{x\to 0} \frac{\alpha(1-e^{-\alpha x})}{2x} - \frac{1 - e^{-\alpha x} - x\alpha e^{-\alpha x}}{x^2} \\ &= \lim_{x\to 0} \frac{\alpha^2}{2} - \frac{\alpha e^{-\alpha x} - \alpha e^{-\alpha x} + \alpha^2 x e^{-\alpha x}}{2x} \\ &= 0, \end{aligned} \tag{A.6.2}$$

where the third equality holds because of L'Hospital's rule.

Then to analyze the monotonicity of $k_1(x)$, we make a simplification as follows: let $p_1(x) := k_1'(x) = (-\alpha^2 x^2 - 3\alpha x - 4)e^{-\alpha x} - \alpha x + 4$. Thus we have $p_1'(x) = \alpha(\alpha^2 x^2 e^{-\alpha x} + \alpha x e^{-\alpha x} + e^{-\alpha x} - 1)$ and $p_1''(x) = \alpha^3 x e^{-\alpha x}(1 - \alpha x)$. Since $x \in (0,1)$, and $\alpha > 0$, then $\alpha^3 x e^{-\alpha x} > 0$. The sign of $1 - \alpha x$ is unascertainable because $\alpha$ can be larger or smaller than 1. When $x \to 0$, a bounded parameter $\alpha$ can assure that $p_1''(x) > 0$, while if $\alpha$ or $x$ is large, $p_1''(x)$ change to negative. Thus there are two probabilities for the sign of $p_1''(x)$ when $x \in (0,1)$: either positive or positive to negative. As a result, the monotonicity of $p_1'(x)$ is either increasing or increasing to decreasing. The boundary conditions of $p_1'(x)$ are:

$$\begin{cases} p_1'(0) = 0 \\ p_1'(1) = \alpha^2 x^2 e^{-\alpha x} + \alpha x e^{-\alpha x} + e^{-\alpha x} - 1 \end{cases} \quad \text{(A.6.3)}$$

When $p_1'(1) = 0$, $\bar{\alpha} = 1.793$. The two cases of the monotonicity of $p_1'(x)$ are shows respectively:

(a) When $\alpha \in (0, \bar{\alpha}]$, $p_1'(x) \geq 0$. $p_1(x)$ is increasing in $x$. $p_1(x) > p_1(0) = 0$. Based on the definition of $k_1'(x)$ and $p_1(x)$, we can conclude $k_1(x)$ is increasing, thus $k_1(x) > k_1(0) = 0$.

(b) When $\alpha > \bar{\alpha}$, the sign of $p_1'(x)$ transforms from positive to negative, therefore, the monotonicity of $p_1(x)$ transforms from increasing to decreasing. We can conclude that: $p_1(x) > \min\{p_1(0), p_1(1)\} = \min\{0, (-\alpha^2 - 3\alpha - 4)e^{-\alpha} - \alpha + 4\}$. There are two cases as follows:

When $p_1(1) = 0$, $\bar{\bar{\alpha}} = 2.688$.

(i) When $\bar{\alpha} < \alpha \leq \bar{\bar{\alpha}}$, $p_1(1) \geq 0$, $p_1(x) > 0$. We can conclude $k_1'(x) > 0$, and $k_1(x)$ is increasing similarly.

(ii) When $\alpha > \bar{\bar{\alpha}}$, the sign of $p_1(x)$ transform positive to negative. According to definition of $k_1'(x)$ and $p_1(x)$, it is obvious that the sign of $k_1'(x)$ is transform positive to negative, and the monotonicity of $k_1(x)$ changes from increasing to decreasing when $x \in (0,1)$. Therefore, we have $k_1(x) > \min\{k_1(0), k_1(1)\}$. Meanwhile, the boundary condition is $k_1(1) = \frac{\alpha}{2} - \frac{\alpha e^{-\alpha}}{2} - 1 + e^{-\alpha} + \alpha e^{-\alpha} = \frac{\alpha e^{-\alpha}}{2} + e^{-\alpha} + \left(\frac{\alpha}{2} - 1\right) > 0$. In conclusion, $k_1(x) > 0$.

(2) Secondly, we consider $k_2(x)$:

The boundary condition is:

$$\lim_{x \to 1} k_2(x) = \lim_{x \to 1} -\frac{\alpha(e^{-\alpha x} - e^{-\alpha})}{2(1-x)} + \alpha e^{-\alpha x} - \frac{\frac{e^{-\alpha x} - e^{-\alpha}}{1-x}}{1-x}$$

Let $u = 1 - x$, and $x = 1 - u$.

$$\lim_{x \to 1} k_2(x) = \lim_{u \to 0} -\frac{\alpha(e^{-\alpha(1-u)} - e^{-\alpha})}{2u} + \frac{\alpha u e^{-\alpha x} - e^{-\alpha(1-u)} + e^{-\alpha}}{u^2}$$

$$= \lim_{u \to 0} e^{-\alpha}(-\frac{\alpha(e^{\alpha u} - 1)}{2u} + \frac{\alpha u e^{\alpha u} - e^{\alpha u} + 1}{u^2})$$

Based on L'Hospital's rule, we have:

$$\lim_{x \to 1} k_2(x) = \lim_{u \to 0} e^{-\alpha}\left(-\frac{\alpha(e^{\alpha u} - 1)}{2u} + \frac{\alpha u e^{\alpha u} - e^{\alpha u} + 1}{u^2}\right)$$

$$= \lim_{u \to 0} e^{-\alpha}\left(-\frac{\alpha^2}{2} + \frac{\alpha e^{\alpha u} + \alpha^2 u e^{\alpha u} - \alpha e^{\alpha u}}{2u}\right)$$

$$= \lim_{u \to 0} e^{-\alpha}\left(-\frac{\alpha^2}{2} + \frac{\alpha^2 u e^{\alpha u}}{2u}\right)$$

$$= 0$$

Then we analyze the monotonicity of $k_2(x)$: first, we make a definition as follows: let the derivative of $k_2(x)$ is $p_2(x) = (4 + \alpha^2(1-x)^2 - 3\alpha(1-x))e^{-\alpha x} - e^{-\alpha}(4 + \alpha(1-x))$, thus $p_2'(x) = -\alpha((1 + \alpha^2(1-x)^2 - \alpha(1-x))e^{-\alpha x} - e^{-\alpha})$, and $p_2''(x) = \alpha(1 + \alpha(1-x))(1-x)\alpha e^{-\alpha x} > 0$. Therefore, $p_2'(x)$ is increasing, and $p_2'(x) < p_2'(1) = 0$. When $x \in (0,1)$, $p_2(x) = (4 + \alpha^2(1-x)^2 - 3\alpha(1-x))e^{-\alpha x} - e^{-\alpha}(4 + \alpha(1-x)) > 0$. According to definition, $k_2'(x) = \frac{p_2(x)}{-2(1-x)^3} < 0$, and $k_2(x) > k_2(1) = 0$.

In summary, the analysis the function $k(x)$, based on the monotonicity of $k_1(x)$ and $k_2(x)$, we have:

$$\min_{x \in (0,1)} k(x) = \min_{x \in (0,1)} (k_1(x) + k_2(x)) \geq \min_{x \in (0,1)} k_1(x) + \min_{x \in (0,1)} k_2(x) = k_1(0) + k_2(1) > 0$$

As a result, we have $k(x) > 0$ and $k_1(x) > 0$, thus Lemma 6 is proved.

### A.7. Proof of Lemma 7

To prove that the function $z(x)$ decreases in $x$, based on Definition 2, we have: $\frac{dz(x)}{dx} = (\gamma - 1)\frac{\partial z_1(x,\gamma)}{\partial x} - \gamma \frac{\partial z_2(x,\gamma)}{\partial x}$. Thus, to prove Lemma 7 is equivalent to prove

$$(\gamma - 1)\frac{\partial z_1(x,\gamma)}{\partial x} < \gamma \frac{\partial z_2(x,\gamma)}{\partial x}. \tag{A.7.1}$$

Note that based on Definition 2,

$$\frac{\partial z_1(x,\gamma)}{\partial x} = -\alpha^2 e^{-\alpha x}(\varphi_1^\gamma - \varphi_2^\gamma) + \alpha e^{-\alpha x}\gamma(\varphi_1^{\gamma-1}\varphi_1'(x) - \varphi_2^{\gamma-1}\varphi_2'(x))$$

(A.7.2)

$$\begin{aligned}\frac{\partial z_2(x,\gamma)}{\partial x} &= \varphi_2'(x)\varphi_1{}^\gamma + \gamma\varphi_2\varphi_1{}^{\gamma-1}\varphi_1'(x) - \varphi_1'(x)\varphi_2{}^\gamma - \gamma\varphi_1\varphi_2{}^{\gamma-1}\varphi_2'(x) \\ &= (\gamma\varphi_2\varphi_1{}^{\gamma-1} - \varphi_2{}^\gamma)\varphi_1'(x) - (\gamma\varphi_1\varphi_2{}^{\gamma-1} - \varphi_1{}^\gamma)\varphi_2'(x).\end{aligned} \quad (A.7.3)$$

By dividing a positive number $\gamma$ in both sides of the inequality (A.7.1), we have that (A.7.1) is equivalent to

$$\frac{(\gamma-1)}{\gamma}\frac{\partial z_1(x,\gamma)}{\partial x} < \frac{\partial z_2(x,\gamma)}{\partial x}. \quad (A.7.4)$$

According to Equation (A.7.2), the LHS of inequality (A.7.4) can be merged by terms of $\varphi_1'(x)$ and $\varphi_2'(x)$ as follows:

$$\begin{aligned}\frac{(\gamma-1)}{\gamma}\frac{\partial z_1(x,\gamma)}{\partial x} &= \alpha e^{-\alpha x}(\gamma-1)\bigl(\varphi_1{}^{\gamma-1}\varphi_1'(x) - \varphi_2{}^{\gamma-1}\varphi_2'(x)\bigr) - \frac{\gamma-1}{\gamma}\alpha^2 e^{-\alpha x}(\varphi_1{}^\gamma - \varphi_2{}^\gamma) \\ &= (\gamma-1)\alpha e^{-\alpha x}\varphi_1{}^{\gamma-1}\varphi_1'(x) - (\gamma-1)\alpha e^{-\alpha x}\varphi_2{}^{\gamma-1}\varphi_2'(x) - \frac{\gamma-1}{\gamma}\alpha^2 e^{-\alpha x}(\varphi_1{}^\gamma - \varphi_2{}^\gamma).\end{aligned}$$

$$(A.7.5)$$

Based on Lemma 1 and Lemma 2, it is easy to find that $(\gamma-1)\varphi_1{}^{\gamma-1} > 0$, $(\gamma-1)\varphi_2{}^{\gamma-1} > 0$ and $\varphi_1'(x), \varphi_2'(x) < 0$. Thus, Equation (A.7.5) and (A.7.3) can be simplified as follows:

$$\begin{aligned}\frac{(\gamma-1)}{\gamma}\cdot\frac{\partial z_1(x,\gamma)}{\partial x} &= -\Bigl[(\gamma-1)\alpha e^{-\alpha x}\varphi_1{}^{\gamma-1}|\varphi_1'(x)| - (\gamma-1)\alpha e^{-\alpha x}\varphi_2{}^{\gamma-1}|\varphi_2'(x)| + \frac{\gamma-1}{\gamma}\alpha^2 e^{-\alpha x}(\varphi_1{}^\gamma - \varphi_2{}^\gamma)\Bigr] \\ &= -\bigl\{\gamma\alpha e^{-\alpha x}\varphi_1{}^{\gamma-1}|\varphi_1'(x)| - \gamma\alpha e^{-\alpha x}\varphi_2{}^{\gamma-1}|\varphi_2'(x)| \\ &\quad - [\alpha e^{-\alpha x}\varphi_1{}^{\gamma-1}|\varphi_1'(x)| - \alpha e^{-\alpha x}\varphi_2{}^{\gamma-1}|\varphi_2'(x)|] + \frac{\gamma-1}{\gamma}\alpha^2 e^{-\alpha x}(\varphi_1{}^\gamma - \varphi_2{}^\gamma)\bigr\}\end{aligned}$$

$$(A.7.6)$$

and

$$\frac{\partial z_2(x,\gamma)}{\partial x} = -[(\gamma\varphi_2\varphi_1{}^{\gamma-1} - \varphi_2{}^\gamma)|\varphi_1'(x)| - (\gamma\varphi_1\varphi_2{}^{\gamma-1} - \varphi_1{}^\gamma)|\varphi_2'(x)|]. \quad (A.7.7)$$

To prove Equation (A.7.4), according to Equation (A.7.6) and (A.7.7), we just need to prove:

$$\begin{aligned}\gamma\alpha e^{-\alpha x}\varphi_1{}^{\gamma-1}|\varphi_1'(x)| &- \gamma\alpha e^{-\alpha x}\varphi_2{}^{\gamma-1}|\varphi_2'(x)| \\ &- [\alpha e^{-\alpha x}\varphi_1{}^{\gamma-1}|\varphi_1'(x)| - \alpha e^{-\alpha x}\varphi_2{}^{\gamma-1}|\varphi_2'(x)|] + \frac{\gamma-1}{\gamma}\alpha^2 e^{-\alpha x}(\varphi_1{}^\gamma - \varphi_2{}^\gamma) \\ &> (\gamma\varphi_2\varphi_1{}^{\gamma-1} - \varphi_2{}^\gamma)|\varphi_1'(x)| - (\gamma\varphi_1\varphi_2{}^{\gamma-1} - \varphi_1{}^\gamma)|\varphi_2'(x)|.\end{aligned}$$

By combining similar terms, we can rewrite the inequality above as follows:

$$\gamma\alpha e^{-\alpha x}\varphi_1{}^{\gamma-1}|\varphi_1'(x)| - \gamma\alpha e^{-\alpha x}\varphi_2{}^{\gamma-1}|\varphi_2'(x)| - [(\gamma\varphi_2\varphi_1{}^{\gamma-1} - \varphi_2{}^\gamma)|\varphi_1'(x)| - (\gamma\varphi_1\varphi_2{}^{\gamma-1} - \varphi_1{}^\gamma)|\varphi_2'(x)|] + \frac{\gamma-1}{\gamma}\alpha^2 e^{-\alpha x}(\varphi_1{}^\gamma - \varphi_2{}^\gamma) - [\alpha e^{-\alpha x}\varphi_1{}^{\gamma-1}|\varphi_1'(x)| - \alpha e^{-\alpha x}\varphi_2{}^{\gamma-1}|\varphi_2'(x)|] > 0.$$

$$(A.7.8)$$

The sufficient condition of that the inequality (A.7.8) holds is the following two inequalities hold simultaneously.

$$\gamma\alpha e^{-\alpha x}\varphi_1{}^{\gamma-1}|\varphi_1{}'(x)| - \gamma\alpha e^{-\alpha x}\varphi_2{}^{\gamma-1}|\varphi_2{}'(x)|$$
$$> (\gamma\varphi_2\varphi_1{}^{\gamma-1} - \varphi_2{}^{\gamma})|\varphi_1{}'(x)| - (\gamma\varphi_1\varphi_2{}^{\gamma-1} - \varphi_1{}^{\gamma})|\varphi_2{}'(x)|$$
(A.7.9)

$$\frac{\gamma-1}{\gamma}\alpha^2 e^{-\alpha x}(\varphi_1{}^{\gamma} - \varphi_2{}^{\gamma}) > \alpha e^{-\alpha x}\varphi_1{}^{\gamma-1}|\varphi_1{}'(x)| - \alpha e^{-\alpha x}\varphi_2{}^{\gamma-1}|\varphi_2{}'(x)|$$
(A.7.10)

(1) Firstly, we prove the inequality (A.7.9).

**Definition A.1**: *To simplify the expression, we denote*

$$\begin{cases} A_1 = (\gamma\varphi_2\varphi_1{}^{\gamma-1} - \varphi_2{}^{\gamma}) \\ A_2 = (\gamma\varphi_1\varphi_2{}^{\gamma-1} - \varphi_1{}^{\gamma}) \end{cases}$$
(A.7.11)

*and*

$$\begin{cases} B_1 = \alpha e^{-\alpha x}\gamma\varphi_1{}^{\gamma-1} \\ B_2 = \alpha e^{-\alpha x}\gamma\varphi_2{}^{\gamma-1}. \end{cases}$$
(A.7.12)

*Moreover, we denote*

$$\begin{cases} \Delta A \coloneqq A_1 - A_2 = (\gamma\varphi_2 - \varphi_1)\varphi_1{}^{\gamma-1} - (\gamma\varphi_1 - \varphi_2)\varphi_2{}^{\gamma-1} \\ \Delta B \coloneqq B_1 - B_2 = \gamma\alpha e^{-\alpha x}\varphi_1{}^{\gamma-1} - \gamma\alpha e^{-\alpha x}\varphi_2{}^{\gamma-1}. \end{cases}$$
(A.7.13)

Based on Definition A.1, we can rewrite the inequality (A.7.9) as inequality (A.7.14):

$$B_1|\varphi_1{}'(x)| - B_2|\varphi_2{}'(x)| > A_1|\varphi_1{}'(x)| - A_2|\varphi_2{}'(x)|. \quad (A.7.14)$$

According to Equations (A.1.2) and (A.2.2), we have $\varphi_1(x) > |f'(x)| > \varphi_2(x) > 0$, and $\gamma > 0$. Thus, we have

$$\begin{cases} \gamma\varphi_2 - \varphi_1 < \gamma|f'(x)| - \varphi_1 < (\gamma-1)|f'(x)| \\ \gamma\varphi_1 - \varphi_2 > \gamma|f'(x)| - \varphi_2 > (\gamma-1)|f'(x)| \end{cases}$$

and so

$$\begin{aligned}\Delta A &= (\gamma\varphi_2 - \varphi_1)\varphi_1{}^{\gamma-1} - (\gamma\varphi_1 - \varphi_2)\varphi_2{}^{\gamma-1} \\ &< (\gamma-1)|f'(x)|\varphi_1{}^{\gamma-1} - (\gamma-1)|f'(x)|\varphi_2{}^{\gamma-1} \\ &= (\gamma-1)|f'(x)|(\varphi_1{}^{\gamma-1} - \varphi_2{}^{\gamma-1}). \end{aligned} \quad (A.7.15)$$

Combining with conditions $\varphi_1 > \varphi_2 > 0$ in Lemma 1 and $\varphi_1{}^{\gamma-1} - \varphi_2{}^{\gamma-1} > 0$ in Lemma 5, we have

$$(\gamma-1)|f'(x)|(\varphi_1{}^{\gamma-1} - \varphi_2{}^{\gamma-1}) < \gamma|f'(x)|(\varphi_1{}^{\gamma-1} - \varphi_2{}^{\gamma-1})$$
$$= \gamma|f'(x)|\varphi_1{}^{\gamma-1} - \gamma|f'(x)|\varphi_2{}^{\gamma-1} = B_1 - B_2 = \Delta B.$$
(A.7.16)

Meanwhile, because of $\varphi_1(x) > |f'(x)| > \varphi_2(x)$, we conclude that

$$B_1 = \alpha e^{-\alpha x}\gamma\varphi_1{}^{\gamma-1} = \gamma|f'(x)|\varphi_1{}^{\gamma-1} > \gamma\varphi_2\varphi_1{}^{\gamma-1} > \gamma\varphi_2\varphi_1{}^{\gamma-1} - \varphi_2{}^{\gamma} = A_1. \quad (A.7.17)$$

Thus, when focusing on the inequality (A.7.14), we have:

LHS $= B_1|\varphi_1'(x)| - B_2|\varphi_2'(x)|$
$= A_1|\varphi_1'(x)| - (B_2 - B_1 + A_1)|\varphi_2'(x)| + (B_1 - A_1)(|\varphi_1'(x)| - |\varphi_2'(x)|).$
(A.7.18)

Since $(B_1 - A_1) > 0$ based on (A.7.17), $\varphi_1'(x) < \varphi_2'(x) < 0$ and $|\varphi_1'(x)| - |\varphi_2'(x)| > 0$ based on Lemma 2, Lemma 5 and Equation (A.7.18), we can conclude that

LHS $> A_1|\varphi_1'(x)| - (B_2 - B_1 + A_1)|\varphi_2'(x)| = A_1|\varphi_1'(x)| - (A_1 - \Delta B)|\varphi_2'(x)|.$
(A.7.19)

Furthermore, according to the inequalities (A.7.12) and (A.7.13), we have

$$\begin{aligned}A_1|\varphi_1'(x)| - (A_1 - \Delta B)|\varphi_2'(x)| &> A_1|\varphi_1'(x)| - (A_1 - \Delta A)|\varphi_2'(x)|\\ &= A_1|\varphi_1'(x)| - A_2|\varphi_2'(x)|\\ &= \text{RHS}.\end{aligned}$$
(A.7.20)

In conclusion, based on (A.7.14), (A.7.19), and (A.7.20), we have that (A.7.9) holds.

(2) Secondly, we show that the inequality (A.7.10) holds.

The inequality (A.7.10) can be simplified as follows

$$\frac{\gamma - 1}{\gamma}\alpha^2 e^{-\alpha x}(\varphi_1^\gamma - \varphi_2^\gamma) > \alpha e^{-\alpha x}\varphi_1^{\gamma-1}|\varphi_1'(x)| - \alpha e^{-\alpha x}\varphi_2^{\gamma-1}|\varphi_2'(x)|$$

Because $\alpha e^{-\alpha x} > 0$, we can divided $\alpha e^{-\alpha x}$ in both sides of inequality (A.7.10) as follows

$$\left(\frac{\gamma - 1}{\gamma}\alpha\varphi_1^\gamma - \frac{\gamma - 1}{\gamma}\alpha\varphi_2^\gamma\right) > \varphi_1^{\gamma-1}|\varphi_1'(x)| - \varphi_2^{\gamma-1}|\varphi_2'(x)|.$$

After rearranging the above inequality we have

$$\left(\frac{\gamma-1}{\gamma}\alpha\varphi_1 - |\varphi_1'(x)|\right)\varphi_1^{\gamma-1} > \left(\frac{\gamma-1}{\gamma}\alpha\varphi_2 - |\varphi_2'(x)|\right)\varphi_2^{\gamma-1}. \quad (A.7.21)$$

Because $\varphi_1^{\gamma-1} > \varphi_2^{\gamma-1}$ according to Lemma 5, we have one sufficient condition of inequality (A.7.21) as

$$\begin{cases}\frac{\gamma-1}{\gamma}\alpha\varphi_1 - |\varphi_1'(x)| > \frac{\gamma-1}{\gamma}\alpha\varphi_2 - |\varphi_2'(x)|\\ \frac{\gamma-1}{\gamma}\alpha\varphi_1 - |\varphi_1'(x)| > 0.\end{cases} \quad (A.7.22)$$

Let $\rho = \frac{\gamma-1}{\gamma} \in [\frac{1}{2}, 1)$. Because $\varphi_1 > \varphi_2$ and $\alpha > 0$, we have that if $\rho = \frac{1}{2}$ the inequality (A.7.22) holds, when $\rho > \frac{1}{2}$ inequality (A.7.22) also holds. Thus, it suffices to prove (A.7.22) when $\rho = \frac{\gamma-1}{\gamma} = \frac{1}{2}$. Based on Definition 1 and Equation (A.2.1) in Lemma 2, we can transform the inequality (A.7.22) to

$$\begin{cases} \dfrac{\alpha}{2}\dfrac{1-e^{-\alpha x}}{x} - \dfrac{\frac{1-e^{-\alpha x}}{x}-\alpha e^{-\alpha x}}{x} > \dfrac{\alpha}{2}\dfrac{e^{-\alpha x}-e^{-\alpha}}{1-x} - \dfrac{\alpha e^{-\alpha x}-\frac{e^{-\alpha x}-e^{-\alpha}}{1-x}}{1-x} \\ \dfrac{\alpha}{2}\dfrac{1-e^{-\alpha x}}{x} - \dfrac{\frac{1-e^{-\alpha x}}{x}-\alpha e^{-\alpha x}}{x} > 0, \end{cases} \quad (A.7.23)$$

which holds based on Lemma 6.

## A.8. Proof of Lemma 8

Lemma 7 proposed the monotonicity of $z(x)$: $z(x)$ is a decreasing function in $x$. We propose the boundary conditions of $z(x)$ in this part.

Firstly, the boundary limitation of $\varphi_1(x)$ and $\varphi_2(x)$ is:

$$\begin{cases} \lim_{x\to 0}\varphi_1(x) = -f'(0) = \alpha \\ \lim_{x\to 0}\varphi_2(x) = (1-e^{-\alpha}) \\ \lim_{x\to 1}\varphi_1(x) = (1-e^{-\alpha}) \\ \lim_{x\to 1}\varphi_2(x) = -f'(0) = \alpha e^{-\alpha} \end{cases} \quad (A.8.1)$$

Next, we show the boundary condition of $z(0)$:

$$\begin{aligned} z(0) &= (\gamma-1)\alpha(\alpha^\gamma - (1-e^{-\alpha})^\gamma) - \gamma\big((1-e^{-\alpha})\alpha^\gamma - \alpha(1-e^{-\alpha})^\gamma\big) \\ &= (\gamma-1)\alpha^{\gamma+1} + \alpha(1-e^{-\alpha})^\gamma - \gamma(1-e^{-\alpha})\alpha^\gamma \\ &= \alpha^{\gamma+1}\Big((\gamma-1) + \Big(\dfrac{1-e^{-\alpha}}{\alpha}\Big)^\gamma - \gamma\cdot\dfrac{1-e^{-\alpha}}{\alpha}\Big) \end{aligned}$$
(A.8.2)

Let $v = \dfrac{1-e^{-\alpha}}{\alpha}$. Based on Lemma 1, we have the condition $\varphi_1(x) > \varphi_2(x)$. And then $v = \dfrac{1-e^{-\alpha}}{\alpha} = \dfrac{\lim_{x\to 0}\varphi_2(x)}{\lim_{x\to 0}\varphi_1(x)} < 1$. Equation (A.8.2) is simplified as: $z(0,\gamma) = \alpha^{\gamma+1}\big((\gamma-1) + v^\gamma - \gamma\cdot v\big) := w(v)$. Because $\dfrac{dw(v)}{dv} = \alpha^{\gamma+1}\gamma(v^{\gamma-1}-1) < 0$, $w(v) > w(1) = 0$. As a result, $z(0) > 0$.

Then we show $z(1,\gamma) < 0$ similarly:

$$\begin{aligned} z(1) &= (\gamma-1)\alpha e^{-\alpha}((1-e^{-\alpha})^\gamma - (\alpha e^{-\alpha})^\gamma) \\ &\quad - \gamma(\alpha e^{-\alpha}(1-e^{-\alpha})^\gamma - \alpha e^{-\alpha}(\alpha e^{-\alpha})^\gamma) \\ &= -(\gamma-1)(\alpha e^{-\alpha})^{\gamma+1} - \alpha e^{-\alpha}(1-e^{-\alpha})^\gamma + \gamma(\alpha e^{-\alpha})^{\gamma+1} \\ &= (\alpha e^{-\alpha})^{\gamma+1} - \alpha e^{-\alpha}(1-e^{-\alpha})^\gamma = \alpha e^{-\alpha}((\alpha e^{-\alpha})^\gamma - (1-e^{-\alpha})^\gamma) \\ &< 0 \end{aligned}$$

In conclusion, we show that $z(x)$ is decreasing function when $x \in (0,1)$, and changed from positive to negative. Thus, $z(x) = 0$ has a unique solution when $x \in$ (0,1). Lemma 8 is proved.

**Appendix B. Proof of Theorem 1**

Given $\theta = \theta_0$, the original optimization problem $\max_{p_1 \geq 0, p_2 \geq 0} \pi(p_1, p_2, \theta_0)$ can be separated into two optimization problems: $\max_{p_1 \geq 0} \pi_1(p_1, \theta_0)$ and $\max_{p_2 \geq 0} \pi_2(p_2, \theta_0)$. Evidently, solving the original optimization problem is equivalent to solving these two sub problems. To demonstrate the uniqueness and optimality of the solutions for the two sub problems, we first compute the first partial derivatives of $p_1$ and $p_2$, as follows:

$$\begin{cases} \dfrac{\partial \pi(p_1, p_2, \theta_0)}{\partial p_1} = \dfrac{d\pi_1(p_1, \theta_0)}{dp_1} = \dfrac{c \cdot \gamma(1 - e^{-\alpha\theta_0}) - \alpha\theta_0(\gamma - 1)p_1}{\alpha p_1^{\gamma+1}} \\ \dfrac{\partial \pi(p_1, p_2, \theta_0)}{\partial p_2} = \dfrac{d\pi_2(p_2, \theta_0)}{dp_2} = \dfrac{c \cdot \gamma(e^{-\alpha\theta_0} - e^{-\alpha}) - \alpha(1 - \theta_0)(\gamma - 1)p_2}{\alpha p_2^{\gamma+1}} \end{cases}.$$

By equating these two equations to 0, and we can obtain the unique solutions as follows: $p_1^*(\theta_0) = \dfrac{c \cdot \gamma(1 - e^{-\alpha\theta_0})}{\alpha\theta_0(\gamma - 1)}$ and $p_2^*(\theta_0) = \dfrac{c \cdot \gamma(e^{-\alpha\theta_0} - e^{-\alpha})}{\alpha(1 - \theta_0)(\gamma - 1)}$. When $p_1(\theta_0) < p_1^*(\theta_0)$, because $\alpha$, $\theta_0$ and $(\gamma - 1)$ are positive, we have $\dfrac{d\pi_1(p_1, \theta_0)}{dp_1} > 0$. Otherwise, when $p_1(\theta_0) > p_1^*(\theta_0)$, $\dfrac{d\pi_1(p_1, \theta_0)}{dp_1} < 0$. Therefore, $p_1^*(\theta_0)$ is the global maximizer of $\pi_1(p_1, \theta_0)$. Similarly, we can prove that $p_2^*(\theta_0)$ is the global maximizer of $\pi_2(p_2, \theta_0)$.

Moreover, according to Definition 1, it is obvious that $p_1^*(\theta_0) = \dfrac{c \cdot \gamma}{\alpha(\gamma - 1)} \varphi_1(\theta_0)$ and $p_2^*(\theta_0) = \dfrac{c \cdot \gamma}{\alpha(\gamma - 1)} \varphi_2(\theta_0)$. According to Lemma 1, $\varphi_1(\theta_0) > \varphi_2(\theta_0)$ when $\alpha > 0$ for any $\theta_0 \in (0,1)$. Because $\gamma$, $c$ and $\alpha$ are positive, we have $p_1 > p_2$ for any $\theta \in (0,1)$. When $\alpha = 0$, we have $p_1^*(\theta_0) = \lim_{\alpha \to 0} \dfrac{c \cdot \gamma(1 - e^{-\alpha\theta_0})}{\alpha\theta_0(\gamma - 1)} = \dfrac{c \cdot \gamma}{\gamma - 1}$ according to L'Hospital's rule. Similarly, we have $\varphi_2(\theta_0) = \lim_{\alpha \to 0} \dfrac{c \cdot \gamma(e^{-\alpha\theta_0} - e^{-\alpha})}{\alpha(1 - \theta_0)(\gamma - 1)} = \dfrac{c \cdot \gamma \cdot e^{-\alpha}(e^{1-\theta_0} - 1)}{\alpha(1 - \theta_0)(\gamma - 1)} = \dfrac{c \cdot \gamma}{\gamma - 1} = \varphi_1(\theta_0)$.

**Appendix C. Computing the Profits based on optimal prices in basic model**

Based on Theorem 1, we have the expression of the optimal prices in two stages:

$p_1^*(\theta_0) = \frac{c \cdot \gamma(1-e^{-\alpha\theta_0})}{\alpha\theta_0(\gamma-1)}$ and $p_2^*(\theta_0) = \frac{c \cdot \gamma(e^{-\alpha\theta_0}-e^{-\alpha})}{\alpha(1-\theta_0)(\gamma-1)}$. Substituting them into the profits expression, we have:

$$\pi(\theta) = \frac{1}{p_1^*(\theta_0)^\gamma} \int_0^{\theta_0} \lambda(t)[p_1^*(\theta_0) - c(t)]dt + \frac{1}{p_2^*(\theta_0)^\gamma} \int_{\theta_0}^1 \lambda(t)[p_2^*(\theta_0) - c(t)]dt.$$

Based on the assumption that $\lambda(t)$ is uniform distribution we have

$$\pi(\theta_0) = \frac{p_1^*(\theta_0) \cdot \theta_0 - \int_0^{\theta_0} c \cdot e^{-\alpha t} dt}{p_1^*(\theta_0)^\gamma} + \frac{p_2^*(\theta_0) \cdot \theta_0 - \int_{\theta_0}^1 c \cdot e^{-\alpha t} dt}{p_2^*(\theta_0)^\gamma}.$$

After simplifying the integral expression $c$, we have

$$\pi(\theta_0) = \frac{p_1^*(\theta_0) \cdot \theta_0 - \frac{c(1-e^{-\alpha\theta_0})}{\alpha}}{p_1^*(\theta_0)^\gamma} + \frac{p_2^*(\theta_0) \cdot \theta_0 - \frac{c(e^{-\alpha\theta_0}-e^{-\alpha})}{\alpha}}{p_2^*(\theta_0)^\gamma}$$

Then we consider the expression of $p_1^*(\theta_0)$ and $p_2^*(\theta_0)$, the total profit expression is:

$$\pi(\theta_0) = \frac{\frac{c \cdot \gamma(1-e^{-\alpha\theta_0})}{\alpha(\gamma-1)} - \frac{c(1-e^{-\alpha\theta_0})(\gamma-1)}{\alpha(\gamma-1)}}{[\frac{c \cdot \gamma(1-e^{-\alpha\theta_0})}{\alpha\theta_0(\gamma-1)}]^\gamma}$$

$$+ \frac{\frac{c \cdot \gamma(e^{-\alpha\theta_0}-e^{-\alpha})}{\alpha(\gamma-1)} - \frac{c(e^{-\alpha\theta_0}-e^{-\alpha})(\gamma-1)}{\alpha(\gamma-1)}}{[\frac{c \cdot \gamma(e^{-\alpha\theta_0}-e^{-\alpha})}{\alpha(1-\theta_0)(\gamma-1)}]^\gamma}.$$

And then we merge the similar terms and simplify the total profit expression

$$\pi(\theta_0) = \frac{\frac{c(1-e^{-\alpha\theta_0})}{\alpha(\gamma-1)}}{[\frac{c \cdot \gamma(1-e^{-\alpha\theta_0})}{\alpha\theta_0(\gamma-1)}]^\gamma} + \frac{\frac{c \cdot (e^{-\alpha\theta_0}-e^{-\alpha})}{\alpha(\gamma-1)}}{[\frac{c \cdot \gamma(e^{-\alpha\theta_0}-e^{-\alpha})}{\alpha(1-\theta_0)(\gamma-1)}]^\gamma}$$

When extracting the coefficients of the polynomial, we have

$$\pi(\theta_0) = \frac{[\alpha(\gamma-1)]^{\gamma-1}}{c^{\gamma-1}\gamma^\gamma} \left[ \frac{\theta_0^\gamma}{(1-e^{-\alpha\theta_0})^{\gamma-1}} + \frac{(1-\theta_0)^\gamma}{(e^{-\alpha\theta_0}-e^{-\alpha})^{\gamma-1}} \right].$$

**Appendix D. Proof of Theorem 2**

Based on Equation (3.7), the derivative of $\pi_0(x)$ is

$$\frac{d\pi_0(x)}{dx} = \frac{x^{\gamma-1}}{(1-e^{-\alpha x})^\gamma}(\gamma(1-e^{-\alpha x}) - x(\gamma-1)\alpha e^{-\alpha x})$$
$$- \frac{(1-x)^{\gamma-1}}{(e^{-\alpha x}-e^{-\alpha})^\gamma}(\gamma(e^{-\alpha x}-e^{-\alpha}) - (1-x)(\gamma-1)\alpha e^{-\alpha x}) =$$
$$= \frac{\alpha^{\gamma-1}x^{\gamma-1}}{(1-e^{-\alpha x})^\gamma}(\gamma(1-e^{-\alpha x}) - x(\gamma-1)\alpha e^{-\alpha x})$$
$$- \frac{\alpha^{\gamma-1}(1-x)^{\gamma-1}}{(e^{-\alpha x}-e^{-\alpha})^\gamma}(\gamma(e^{-\alpha x}-e^{-\alpha}) - (1-x)(\gamma-1)\alpha e^{-\alpha x})$$

(D.1)

Based on Definition 1, Equation (D.1) can be simplified as

$$\frac{d\pi_0(x)}{dx} = \gamma\left(\frac{1}{\varphi_1^{\gamma-1}} - \frac{1}{\varphi_2^{\gamma-1}}\right) - (\gamma-1)\alpha e^{-\alpha x}\left(\frac{1}{\varphi_1^\gamma} - \frac{1}{\varphi_2^\gamma}\right)$$
$$= \varphi_1^{-\gamma}\varphi_2^{-\gamma} \cdot \left(\gamma(\varphi_1\varphi_2^\gamma - \varphi_2\varphi_1^\gamma) - (\gamma-1)\alpha e^{-\alpha x}(\varphi_2^\gamma - \varphi_1^\gamma)\right)$$
$$= \varphi_1^{-\gamma}\varphi_2^{-\gamma} \cdot \left((\gamma-1)\alpha e^{-\alpha x}(\varphi_1^\gamma - \varphi_2^\gamma) - \gamma(\varphi_2\varphi_1^\gamma - \varphi_1\varphi_2^\gamma)\right)$$

Furthermore, we can simplify $\frac{d\pi_0(x)}{dx}$ based on Definition 2:

$$\frac{d\pi_0(x)}{dx} = \varphi_1^{-\gamma}\varphi_2^{-\gamma} \cdot z(x) \tag{D.2}$$

According to Lemma 1, we have $\varphi_1^{-\gamma}(\theta)\varphi_2^{-\gamma}(\theta) > 0$. Combined with Definition 2, we conclude that the signs of $\frac{d\pi_0(x)}{dx}$ and $\frac{d\pi(x)}{dx}$ are consistent with $z(x)$. The unique zero point of $z(x)$ is also the optimal solution of $\pi_0(x)$ and $\pi(x)$. According to Lemmas 7 and 8, there exists a unique change point for the sign of $z(x)$ from positive to negative. Consequently, we conclude that $\pi_0(x)$ and $\pi(x)$ have a unique maximizer.

**Appendix E. Proof of Proposition 1**

According to the Equation (3.8), we have the expression of $p_1$ and $p_2$:

$$\begin{cases} p_1 = \frac{c \cdot \gamma}{\alpha(\gamma-1)} \cdot \frac{N\left(1-e^{-\frac{\alpha M}{N}}\right) + (r_0-r_m)\frac{\left(1-\beta^M e^{-\frac{\alpha M}{N}}\right)}{1-\beta}}{M + (r_0-r_m)\frac{1-\beta^M}{1-\beta}} \\ p_2 = \frac{c \cdot \gamma}{\alpha(\gamma-1)} \cdot \frac{N(e^{-\frac{\alpha M}{N}} - e^{-\alpha}) + (r_0-r_m)\frac{\beta^M(e^{-\frac{\alpha M}{N}} - \beta^{N-M}e^{-\alpha})}{1-\beta}}{(N-M) + (r_0-r_m)\frac{\beta^M(1-\beta^{N-M})}{1-\beta}} \end{cases} \tag{E.1}$$

Because $\frac{c \cdot \gamma}{\alpha(\gamma-1)} > 0$, to prove $p_1 > p_2$, we just have to show that

$$\frac{N\left(1-e^{-\frac{\alpha M}{N}}\right)+(r_0-r_m)\frac{\left(1-\beta^M e^{-\frac{\alpha M}{N}}\right)}{1-\beta}}{M+(r_0-r_m)\frac{1-\beta^M}{1-\beta}} > \frac{N(e^{-\frac{\alpha M}{N}}-e^{-\alpha})+(r_0-r_m)\frac{\beta^M(e^{-\frac{\alpha M}{N}}-\beta^{N-M}e^{-\alpha})}{1-\beta}}{(N-M)+(r_0-r_m)\frac{\beta^M(1-\beta^{N-M})}{1-\beta}}.$$

**Definition 3:** *Let* $\begin{cases} \frac{C_1}{D_1} = \frac{N\left(1-e^{-\frac{\alpha M}{N}}\right)}{M} \\ \frac{C_2}{D_2} = \frac{N(e^{-\frac{\alpha M}{N}}-e^{-\alpha})}{N-M} \end{cases}$ *and* $\begin{cases} \frac{E_1}{F_1} = \frac{(r_0-r_m)\frac{\left(1-\beta^M e^{-\frac{\alpha M}{N}}\right)}{1-\beta}}{(r_0-r_m)\frac{1-\beta^M}{1-\beta}} = \frac{1-\beta^M e^{-\frac{\alpha M}{N}}}{1-\beta^M} \\ \frac{E_2}{F_2} = \frac{(r_0-r_m)\frac{\beta^M(e^{-\frac{\alpha M}{N}}-\beta^{N-M}e^{-\alpha})}{1-\beta}}{(r_0-r_m)\frac{\beta^M(1-\beta^{N-M})}{1-\beta}} = \frac{e^{-\frac{\alpha M}{N}}-\beta^{N-M}e^{-\alpha}}{1-\beta^{N-M}} \end{cases}$

Define $\theta = \frac{M}{N}$, $\frac{C_1}{D_1} = \frac{1-e^{-\alpha\theta}}{\theta} = \varphi_1(\theta)$ and $\frac{C_2}{D_2} = \frac{e^{-\alpha\theta}-e^{-\alpha}}{1-\theta} = \varphi_2(\theta)$, according to Lemma 1, we have $\frac{C_1}{D_1} > \frac{C_2}{D_2}$. For $\frac{E_1}{F_1} = e^{-\frac{\alpha M}{N}} \cdot \frac{1-\beta^M e^{-\frac{\alpha M}{N}}}{1-\beta^M} = e^{-\frac{\alpha M}{N}} \cdot \frac{1-(\beta e^{-\frac{\alpha}{N}})^M}{1-\beta^M}$, because $\beta \in (0,1)$ and $e^{-\frac{\alpha}{N}} \in (0,1)$, thus $\beta e^{-\frac{\alpha}{N}} < \beta$, and $\frac{E_1}{F_1} > 1$. Meanwhile, $\frac{E_2}{F_2} = e^{-\frac{\alpha M}{N}} \cdot \frac{1-(\beta e^{-\frac{\alpha}{N}})^{N-M}}{1-\beta^{N-M}}$, the smaller $M$ is, the large $N-M$ is and the larger $\frac{1-(\beta e^{-\frac{\alpha}{N}})^{N-M}}{1-\beta^{N-M}}$ is. Moreover, $e^{-\frac{\alpha M}{N}}$ also decreases with $M$. We have $\frac{E_2}{F_2} = e^{-\frac{\alpha M}{N}} \cdot \frac{1-(\beta e^{-\frac{\alpha}{N}})^{N-M}}{1-\beta^{N-M}} < 1$, thus $\frac{E_1}{F_1} > \frac{E_2}{F_2}$.

Therefore, we can conclude $\frac{C_1+E_1}{D_1+F_1} > \frac{C_2+E_2}{D_2+F_2}$, which donates that

$$\frac{N\left(1-e^{-\frac{\alpha M}{N}}\right)+(r_0-r_m)\frac{\left(1-\beta^M e^{-\frac{\alpha M}{N}}\right)}{1-\beta}}{M+(r_0-r_m)\frac{1-\beta^M}{1-\beta}} > \frac{N(e^{-\frac{\alpha M}{N}}-e^{-\alpha})+(r_0-r_m)\frac{\beta^M(e^{-\frac{\alpha M}{N}}-\beta^{N-M}e^{-\alpha})}{1-\beta}}{(N-M)+(r_0-r_m)\frac{\beta^M(1-\beta^{N-M})}{1-\beta}},$$ and $p_1 > p_2$.

**Appendix F. Proof of Theorem 3**

When the total number of periods $N$ is large, and the interval length of each period tends to zero, because $\beta$ is strictly less than 1, it is obvious that $\beta^M$ and $\beta^{N-M}$ approach to zero. We can simplify Equation (3.9) as follows:

$$\begin{cases} \lim\limits_{N\to\infty} \pi_1 = \frac{[\alpha(\gamma-1)]^{\gamma-1}}{\gamma^\gamma c^{\gamma-1}} \cdot \frac{\left(\frac{M}{N}\right)^\gamma}{\left(1-e^{-\alpha\frac{M}{N}}\right)^{\gamma-1}} \\ \lim\limits_{N\to\infty} \pi_2 = \frac{[\alpha(\gamma-1)]^{\gamma-1}}{\gamma^\gamma c^{\gamma-1}} \cdot \frac{\left(\frac{N-M}{N}\right)^\gamma}{\left(e^{-\alpha\frac{M}{N}}-e^{-\alpha}\right)^{\gamma-1}}. \end{cases} \quad (F.1)$$

For the sake of simplicity, let $\theta = \frac{M}{N}$; the expression for the total profit is as follows:

$$\pi(\theta) = \pi_1(\theta) + \pi_2(\theta) = \frac{[\alpha(\gamma-1)]^{\gamma-1}}{\gamma^\gamma c^{\gamma-1}} \cdot \left[\frac{\theta^\gamma}{(1-e^{-\alpha\theta})^{\gamma-1}} + \frac{(1-\theta)^\gamma}{(e^{-\alpha\theta}-e^{-\alpha})^{\gamma-1}}\right],$$

which is consistent with Equation (3.6) in the basic model. The optimality and uniqueness pertaining to Equation (3.6) has been proved in Theorem 2. In practice, if the total learning period of the WOM tends to infinity, indicating that the WOM learning frequency of the customers is as large as possible, the customers are almost aware of the real reliability of the products, and the fluctuation in the sales volume is nonexistent.

**Appendix G. Proof of Theorem 4**

Based on the Equation (3.9) and (F.1), we have the total profit in basic model is

$$\begin{cases} \lim_{N \to \infty} \pi_1 = \frac{[\alpha(\gamma-1)]^{\gamma-1}}{\gamma^\gamma c^{\gamma-1}} \cdot \frac{\left(\frac{M}{N}\right)^\gamma}{\left(1-e^{-\alpha\frac{M}{N}}\right)^{\gamma-1}} \\ \lim_{N \to \infty} \pi_2 = \frac{[\alpha(\gamma-1)]^{\gamma-1}}{\gamma^\gamma c^{\gamma-1}} \cdot \frac{\left(\frac{N-M}{N}\right)^\gamma}{(e^{-\alpha\frac{M}{N}}-e^{-\alpha})^{\gamma-1}} \end{cases}$$ and in reliability learning model by WOM is:

$$\begin{cases} \lim_{N \to \infty} \pi_1 = \frac{[\alpha(\gamma-1)]^{\gamma-1}}{\gamma^\gamma c^{\gamma-1}} \cdot \frac{\left(\frac{M}{N}+\frac{(1-\beta^M)(r_0-r_m)}{N(1-\beta)}\right)^\gamma}{\left((1-e^{-\alpha\frac{M}{N}})+\frac{\left(1-\beta^M e^{-\frac{\alpha M}{N}}\right)(r_0-r_m)}{N(1-\beta)}\right)^{\gamma-1}} \\ \lim_{N \to \infty} \pi_2 = \frac{[\alpha(\gamma-1)]^{\gamma-1}}{\gamma^\gamma c^{\gamma-1}} \cdot \frac{\left(\frac{N-M}{N}+\frac{\beta^M(1-\beta^{N-M})(r_0-r_m)}{N(1-\beta)}\right)^\gamma}{\left(\left(e^{-\frac{\alpha M}{N}}-e^{-\alpha}\right)+\frac{\beta^M\left(e^{-\frac{\alpha M}{N}}-\beta^{N-M}e^{-\alpha}\right)(r_0-r_m)}{N(1-\beta)}\right)^{\gamma-1}} \end{cases}.$$

Taking $\pi_1$ as an example, because the coefficient is same, we just have to compare

$$\frac{\left(\frac{M}{N}\right)^\gamma}{\left(1-e^{-\alpha\frac{M}{N}}\right)^{\gamma-1}} \text{ and } \frac{\left(\frac{M}{N}+\frac{(1-\beta^M)(r_0-r_m)}{N(1-\beta)}\right)^\gamma}{\left((1-e^{-\alpha\frac{M}{N}})+\frac{\left(1-\beta^M e^{-\frac{\alpha M}{N}}\right)(r_0-r_m)}{N(1-\beta)}\right)^{\gamma-1}}.$$

Based on the definition in Theorem 4, we just have to compare $\frac{A_1^\gamma}{B_1^{\gamma-1}}$ and $\frac{(A_1+C_1)^\gamma}{(B_1+D_1)^{\gamma-1}}$. If the profits of the model considering the WOM is higher than benchmark, we have $\frac{A_1^\gamma}{B_1^{\gamma-1}} < \frac{(A_1+C_1)^\gamma}{(B_1+D_1)^{\gamma-1}}$, which is equivalent to $A_1^\gamma(B_1+D_1)^{\gamma-1} < B_1^{\gamma-1}(A_1+C_1)^\gamma$. After transposition, we have $\frac{A_1^\gamma}{(A_1+C_1)^\gamma} < \frac{B_1^{\gamma-1}}{(B_1+D_1)^{\gamma-1}}$, and thus $\frac{A_1}{A_1+C_1} < (\frac{A_1+C_1}{B_1+D_1} \cdot \frac{B_1}{A_1})^{\gamma-1}$.

Similarly for $\pi_2$, the model considering the WOM is higher than benchmark, we also have $\frac{A_2}{A_2+C_2} < (\frac{A_2+C_2}{B_2+D_2} \cdot \frac{B_2}{A_2})^{\gamma-1}$. Therefore, if the total profit considering the WOM is higher than benchmark, the sufficient and necessary condition is $\frac{A_1}{A_1+C_1} + \frac{A_2}{A_2+C_2} < (\frac{A_1+C_1}{B_1+D_1} \cdot \frac{B_1}{A_1})^{\gamma-1} + (\frac{A_2+C_2}{B_2+D_2} \cdot \frac{B_2}{A_2})^{\gamma-1}$.

**Appendix H. Proof of Proposition 2**

According to Theorem 4, the sufficient condition of $\frac{A_1}{A_1+C_1} + \frac{A_2}{A_2+C_2} < (\frac{A_1+C_1}{B_1+D_1} \cdot \frac{B_1}{A_1})^{\gamma-1} + (\frac{A_2+C_2}{B_2+D_2} \cdot \frac{B_2}{A_2})^{\gamma-1}$ is $\begin{cases} \frac{A_1}{A_1+C_1} < (\frac{A_1+C_1}{B_1+D_1} \cdot \frac{B_1}{A_1})^{\gamma-1} \\ \frac{A_2}{A_2+C_2} < (\frac{A_2+C_2}{B_2+D_2} \cdot \frac{B_2}{A_2})^{\gamma-1} \end{cases}$ hold simultaneously.

Taking $\frac{A_1}{A_1+C_1} < (\frac{A_1+C_1}{B_1+D_1} \cdot \frac{B_1}{A_1})^{\gamma-1}$ as an example, we rearrange the terms as $A_1(\frac{A_1}{B_1})^{\gamma-1} < (A_1+C_1)(\frac{A_1+C_1}{B_1+D_1})^{\gamma-1}$. When customers over estimate the reliability, $r_0 - r_m > 0$, $C_1 = \frac{(1-\beta^M)(r_0-r_m)}{N(1-\beta)} > 0$. If we have the condition $\frac{A_1}{B_1} < \frac{C_1}{D_1}$, it is easy to prove that $\frac{A_1+C_1}{B_1+D_1} > \frac{A_1}{B_1}$. Meanwhile, $A_1 + C_1 > A_1$, and $\gamma - 1 > 0$, we can prove $\frac{A_1}{A_1+C_1} < (\frac{A_1+C_1}{B_1+D_1} \cdot \frac{B_1}{A_1})^{\gamma-1}$ holds. Similarly, we can prove $\frac{A_2}{A_2+C_2} < (\frac{A_2+C_2}{B_2+D_2} \cdot \frac{B_2}{A_2})^{\gamma-1}$ by condition $\frac{A_2}{B_2} < \frac{C_2}{D_2}$.

**Appendix I. Robustness analysis of the model based on other distributions:**

In Section 3, we assumed that the probability density function of the demand distribution is uniform over time. However, in practice, as the products' lifecycle progresses, the sales periods can be divided into four stages, namely, the startup, development, mature and recession stages. The sales volume increases to a stable value and later descends to zero until the product exits the market. Thus, in this section, the lifecycle characteristics are considered. Instead of a uniform distribution over time, normal distributions are considered. In most cases, in reality, the demand for products conforms to the normal distribution, and it is thus the most popular configuration to model the demand (Strijbosch et al., 2006).

Although a closed form expression does not exist for the cumulative distribution function of the normal distribution, we can demonstrate that there exists a unique set of optimal solutions to maximize the total profits. Figure I.1(a) illustrates that in the case without warranty (Case II), based on the normal distributed demand, the optimal product prices in the two stages and the switch time can be determined. Figure I.1(b) shows that if we introduce the warranty price to the model (Case IV), the optimal solution to maximize the total profits exists. Thus, the optimality of the proposed model exhibits robustness and is applicable even when the demand does not follow the uniform distribution.

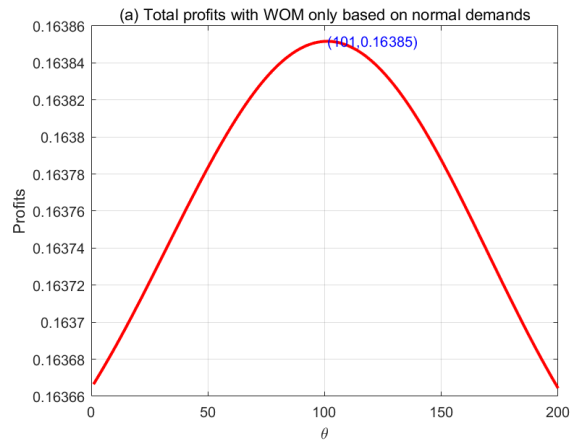

Figure I.1(a) Total profit in the case with WOM only

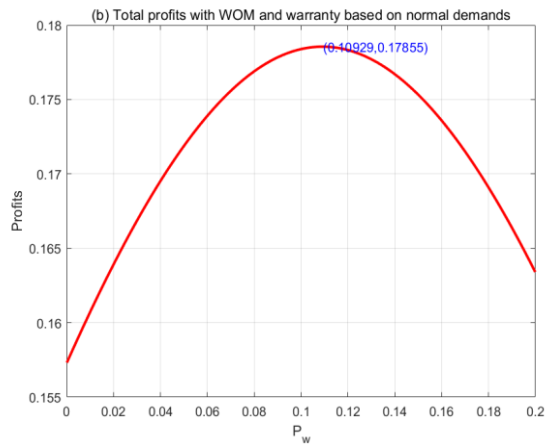

Figure I.1(b) Total profit in the case with the WOM and warranty